\documentclass[preprints,article,accept,moreauthors,pdftex,10pt,a4paper]{Definitions/mdpi}
\pdfoutput=1

\usepackage{booktabs}
\usepackage{multirow}
\usepackage{soul}
\usepackage{microtype}
\graphicspath{{Definitions/}}
\usepackage{environ}
\NewEnviron{myequation}{%
\begin{equation}
\scalebox{0.88}{$\BODY$}
\end{equation}}
\firstpage{1}
\pubvolume{7}
\issuenum{2}
\articlenumber{124}
\pubyear{2019}
\copyrightyear{2019}
\history{Received: 3 November 2018; Accepted: 22 January 2019; Published: 24 January 2019}
\usepackage{longtable}
\usepackage{array}

\usepackage{pdflscape}
\usepackage{bm}
\usepackage{bbold}
\usepackage{amsfonts}
\usepackage{bbm}
\usepackage{rotating}
\usepackage{mathtools}
\usepackage{thmtools,thm-restate}

\renewcommand{\vec}[1]{\underline{#1}}

\newcommand{\dom}[1]{\mathop{dom}(#1)}

\newcommand{\range}[1]{\mathop{range}(#1)}
\newcommand{\supp}[1]{\mathop{supp}(#1)}

\makeatletter
\def\bstr{b}
\def\bfstr{bf}
\def\cstr{c}
\def\fstr{f}
\def\lst{A,B,C,D,d,E,F,G,H,I,J,K,L,M,N,O,P,Q,R,S,T,U,V,W,X,Y,Z,b}

\newcommand{\MkB}[1]{\expandafter\def\csname\bstr#1\endcsname{\mathbb{#1}}}
\@for\i:=\lst\do{%
\expandafter\MkB \i     }

\newcommand{\MkBF}[1]{\expandafter\def\csname\bfstr#1\endcsname{\mathbf{#1}}}
\@for\i:=\lst\do{%
\expandafter\MkBF \i     }

\newcommand{\MkCal}[1]{\expandafter\def\csname\cstr#1\endcsname{\mathcal{#1}}}
\@for\i:=\lst\do{%
\expandafter\MkCal \i     }

\newcommand{\MkFrak}[1]{\expandafter\def\csname\fstr#1\endcsname{\mathfrak{#1}}}
\@for\i:=\lst\do{%
\expandafter\MkFrak \i     }

\makeatother

\newcommand{\LDO}[1]{\mathop{\mathbb{L}_{(#1)}}}

\newcommand{\Mod}[1]{\ \mathrm{mod}\ #1}

\newmuskip\pFqmuskip

\newcommand*\pFq[6][8]{%
\begingroup % only local assignments
\pFqmuskip=#1mu\relax
\mathchardef\normalcomma=\mathcode`,
\mathcode`\,=\string"8000
\begingroup\lccode`\~=`\,
\lowercase{\endgroup\let~}\pFqcomma
{}_{#2}F_{#3}{\left[\genfrac..{0pt}{}{#4}{#5};#6\right]}%
\endgroup
}
\newcommand{\pFqcomma}{{\normalcomma}\mskip\pFqmuskip}

\DeclarePairedDelimiterXPP\seq[2]{}{\big(}{\big)}{_{#2}}{#1}

\newcommand{\tio}{\hat{\mathbf{\mathbb{I}}}}

\newcolumntype{L}{>{$}l<{$}} % math-mode version of "l" column type

\newcolumntype{R}{>{$}r<{$}} % math-mode version of "r" column type

\newcolumntype{C}{>{$}c<{$}} % math-mode version of "c" column type

\Title{Operational Methods in the Study of Sobolev-Jacobi~Polynomials}

\Author{Nicolas Behr $^{1,}$*, Giuseppe Dattoli $^{2}$, G\'{e}rard H. E.~Duchamp $^{3}$, Silvia Licciardi $^{2}$\\ and Karol A.~Penson $^{4}$} 
\AuthorNames{Nicolas Behr, Giuseppe Dattoli, G\'{e}rard H.E.~Duchamp, Silvia Licciardi and Karol A.~Penson}

\address{%
$^{1}$ \quad Institut de Recherche en Informatique Fondamentale (IRIF), Universit\'{e} Paris-Diderot, F-75013 Paris, France\\
$^{2}$ \quad ENEA---Frascati Research Center, Via Enrico Fermi 45, 00044 Rome, Italy; Giuseppe.Dattoli@enea.it (G.D.); silvia.licciardi@dmi.unict.it (S.L.)\\
$^{3}$ \quad Laboratoire d'Informatique de Paris-Nord (LIPN), CNRS UMR 7030, Universit\'{e} Paris 13, Sorbonne Paris Cit\'{e}, F-93430 Villetaneuse, France; ghed@lipn.univ-paris13.fr\\
$^{4}$ \quad Laboratoire de Physique Theorique de la Mati\`{e}re Condens\'{e}e (LPTMC),  CNRS UMR 7600, Sorbonne~Universit\'{e}s, Universit\'{e} Pierre et Marie Curie, F-75005 Paris, France; penson@lptl.jussieu.fr}

\corres{Correspondence: nicolas.behr@irif.fr}

\abstract{Inspired by ideas from umbral calculus and based on the two types of integrals occurring in the defining equations for the gamma and the reciprocal gamma functions, respectively, we~develop a multi-variate version of umbral calculus and of the so-called umbral image technique. Besides~providing a class of new formulae for generalized hypergeometric functions and an implementation of series manipulations for computing lacunary generating functions, our main application of these  techniques is the study of Sobolev-Jacobi polynomials. Motivated by applications to theoretical chemistry, we moreover present a deep link between generalized normal-ordering techniques introduced by Gurappa and Panigrahi, two-variable Hermite polynomials and our integral-based series transforms. Notably, we thus calculate all $K$-tuple $L$-shifted lacunary exponential generating functions for a certain family of Sobolev-Jacobi (SJ) polynomials explicitly.}

\keyword{Sobolev-Jacobi polynomials; umbral image techniques; generalized normal-ordering; lacunary generating functions}

\MSC{33C45; 05A40 (Primary) 30B10; 05A15 (Secondary)}

\begin{document}

\section{Introduction}

The operational methods established in the second half of the 19th century have paved the way to the formalism of quantum mechanics and of umbral calculus~\cite{roman1978umbral}.~The formulation and the technicalities of the quasi-monomiality~\cite{dattoli2000generalized} deepened their roots in the operational formalism and offer a powerful tool to simplify most of the computations associated with the handling of special functions relevant e.g., to the evaluation of generating functions, integrals, and other associated issues.~The~methods which we will exploit in this paper trace back to the original formulation~\cite{roman1978umbral,roman1984umbral}, merged with the complementary aspects developed within the context of the solution of evolution problems~\cite{dattoli1997evolution} and the theory of generalized special functions~\cite{dattoli1999operational}. {Some of the aforementioned complementary aspects were already developed in a wide range of fields, including astrophysics~\cite{rA-Dattoli:2007}, quantum mechanics~\cite{rB-Zhukovsky:2011}, free-electron lasers~\cite{rC-Artioli:2017} and heat conduction~\cite{rD-Dattoli:2017}, with some of the most recent developments reported in~\cite{rE-Zhukovsky:2017,rF-Zhukovsky:2016,rG-Zhukovsky:2016,rH-Zhukovsky:2017,rI-Zhukovsky:2016} (which are based on~\cite{dattoli2000generalized,roman1984umbral,dattoli1997evolution,dattoli1999operational,rJ-Dattoli:2007}).}\\ %

Among the well-established results of such operational methods are formulae such as the operational definition of the two-variable Hermite polynomials $H^{(m)}_n(x,y)$:
\begin{equation}
H_n^{(m)}(x,y):=e^{y\frac{d^m}{dx^m}}x^n =n! \sum_{k=0}^{\lfloor\frac{n}{m}\rfloor}\frac{x^{n-km}y^k}{(n-km)!k!}\,
\end{equation}

In this paper, based on a remarkable operational technique developed by Gurappa and Panigrahi~\cite{gurappa1996new,gurappa2001novel,gurappa2002linear,Gurappa_2003}, we will present an extension of the concept of such exponential formulae by virtue of including a rational function of the Euler operator $D=x\frac{d}{dx}$ in the exponential. This will provide a definition of the Sobolev-Jacobi (SJ) polynomials, which are the central objects of our study~here.

The second main technical tool for our study is the so-called \emph{umbral image} method, which has been introduced in~\cite{babusci2013symbolic,babusci2014spherical,dattoli2017operational} {(see also~\cite{rJ-Dattoli:2007,licciardi2018umbral})}. In favorable cases, this method simplifies computations associated with special functions or orthogonal polynomials considerably. Some concrete examples of such techniques are presented in Section~\ref{sec:ST}. However, in the applications related to SJ polynomials studied in this paper, we will need a certain multi-variate extension of the umbral image method, which we realize in the form of integral transforms related to gamma function-type integrals. This allows us to clarify en passant several technical details that remain somewhat implicit in the traditional umbral calculus approach. Our technique is entirely elementary and, while motivated by and largely analogous to the ideas of umbral calculus, is mathematically self-contained.

{ Referring the interested readers to~\cite{bdp2017} for a more detailed exposition and further technical details, we would like to briefly mention our original motivation in studying SJ polynomials, which stems from the theory of chemical reaction systems.~It is customary in this field to encode the dynamics of such systems in terms of certain probability generating functions and evolution equations thereof, thus in certain special cases, (generalized) orthogonal polynomials may take the role of eigenfunctions of the evolution operators and may thus permit to solve the dynamics problem in closed form. The~simplest possible example in which SJ polynomials arise is the case of a chemical reaction system of one species of molecules and with a single reaction of the form $2X\rightharpoonup\emptyset$ (whence in each transition two particles of type $X$ decay, for simplicity at unit base rate). Mathematically, we may describe the system via defining a probability generating function (with~real-valued time parameter $t$ and formal variable $x$)
\begin{equation}
P(t;x):=\sum_{n\geq 0} p_n(t)x^n\,
\end{equation}
where $x^n$ encodes the system state of having $n$ particles present, and $p_n(t;x)$ the probability of encountering such a state at time $t$. The evolution equation of this reaction system reads
\begin{equation}\label{eq:crEv}
\tfrac{\partial}{\partial} P(t;x)=(1-\hat{x}^2)\tfrac{\partial^2}{\partial x^2}\, P(t;x)\qquad\qquad  (t\geq 0)\,
\end{equation}
with $\hat{x}$ the formal multiplication operator on formal power series (i.e., $\hat{x}(x^n)=x^{n+1}$). As will be presented in full technical detail in Section~\ref{sec:SJP}, the Sobolev-Jacobi polynomials of a particular type are precisely the eigenfunctions of the differential operator $(1-\hat{x}^2)\tfrac{\partial^2}{\partial x^2}$. Given an initial system state, which~could for example consist in a state with precisely $N_0$ particles (whence $P(0;x)=x^{N_0}$), one may provide a closed-form solution to~\eqref{eq:crEv} if it is in addition possible to express the initial configuration in terms of the SJ polynomials, a computation which requires knowledge of the so-called connection coefficients and which is performed in Section~\ref{sec:CCSJ}.~We mention in passing that the Sobolev-Jacobi polynomials for the other set of parameters as discussed in Section~\ref{sec:SJalt} are another particularly interesting case of relevance to chemical reaction systems, since as explained in~\cite{bdp2017} they permit to solve exactly problems involving reaction systems with all three elementary decay-type reactions of one particle species present.}

\textls[-5]{\textbf{Structure of the paper:} In Section~\ref{sec:ST}, we introduce an integral transform technique that permits to implement the analogue of multi-variate umbral images in the spirit of umbral calculus. Sobolev-Jacobi polynomials are introduced in Section~\ref{sec:SJP}, and some first results for a particular subclass of such polynomials are presented in Section~\ref{sec:SJalt}.~The remaining class of Sobolev-Jacobi (SJ) polynomials is studied in detail starting from Section~\ref{sec:SJmain}, including a type of generalized normal-ordering technique involving Euler operators in Section~\ref{sec:ENO} that is of relevance to these considerations.~We then proceed to compute both exponential and lacunary exponential generating functions for the SJ polynomials in Sections~\ref{sec:SJgf}~and~\ref{sec:LGSJ}. This study is made possible due to a discovery of a deep relationship between two-variable Hermite polynomials and SJ polynomials, which is explained in Section~\ref{sec:SJSTT}, and which relies on our series transform techniques (which coincidentally also permit a concise definition of lacunary exponential generating functions (EGF) themselves, see Section~\ref{sec:LGFgen}).~Finally, we devote Section~\ref{sec:CCSJ} to the study of the connection coefficients of the SJ polynomials, and provide some further details and proofs in the technical~appendices}.

\section{Multi-Variate Umbral Calculus via Integral Transforms}
\label{sec:ST}

The key result of this section will be a reformulation (and suitable extension of) the so-called \emph{umbral image method}~\cite{rJ-Dattoli:2007,babusci2013symbolic,babusci2014spherical,dattoli2017operational,babusci2011ramanujan}, which permits to perform certain widely applicable types of series transforms. Rather than working with the umbral calculus notions of operators acting on ``umbral~vacua'' as typical in the more traditional formulations of umbral calculus (see e.g., \cite{licciardi2018umbral} for an extensive overview of modern umbral calculus), we opt here for an alternative approach: series~transforms will be implemented based on certain elementary integrals.~Our motivations are two-fold: firstly, we wish to make our techniques accessible to a wider audience, and thus found it necessary to develop a formulation that is both mathematically sound and firmly rooted in well-established elementary concepts. Secondly, the calculations in this paper will often require the analogues of \emph{multi-variate} umbral calculus techniques, which are immediately accessible in our present formalism, but in tendency would require certain somewhat ad hoc constructions in the traditional approach. For~the interested readers' convenience, we will present several illustrative examples, in~which we will also comment on the relationship to the more traditional umbral calculus approach.

The two types of integrals upon which our calculus will be based are those of the defining equations for the gamma function~(Equation~(5.2.1) in \cite{DLMF}),
\begin{equation}\label{eq:defGamma}
\Gamma(z):=\int_0^{\infty}dt\; e^{-t}t^{z-1}\,
\end{equation}
and of the reciprocal gamma function~(Equation~(5.9.2) in \cite{DLMF})
\begin{equation}\label{eq:recGamma}
\frac{1}{\Gamma(z)}=\frac{1}{2\pi i}\int_{\gamma}dt\; e^{t} t^{-z}\,
\end{equation}
here, according to loc.\ cit., $t^{-z}$ is understood as having its principal value where $t$ crosses the positive real axis and is continuous, and the integration is taken along the \emph{Hankel contour} $\gamma$ (see~\cite{DLMF}, Image~5.9.F1). It is well-known that $\Gamma(z)$ as defined in~\eqref{eq:defGamma} (and for $Re(z)<0$ by analytic continuation) is a meromorphic function with no zeros, and simple poles at $z=-n$ ($n\in \bZ_{\leq 0}$) of residue $(-1)^n/n!$. The reciprocal gamma function $1/\Gamma(z)$ is an entire function with simple zeros at $z=-n$ ($n\in \bZ_{\leq 0}$). The gamma function (resp. reciprocal gamma function) used here will be its full analytic continuation to $\bC\setminus\bZ_{\leq 0}$ (resp. $\bC$).

As a second key ingredient, we will need the following notion of \emph{generalized monomials} (and of formal power series thereover):
\begin{Definition}
Let $\cA$ be an alphabet (i.e., a set of symbols or of formal variables), and let $\bC^{(\cA)}$ denote the set of all functions $\alpha:\cA\rightarrow \bC$ with \emph{finite support},
\begin{equation}
\bC^{(\cA)}:=\{
\alpha:\cA\rightarrow \bC \mid |\supp{\alpha}|<\infty
\}\,
\end{equation}
Here, we define the support $\supp{\alpha}$ of a function $\alpha:\cA\rightarrow \bC$ as the set of elements $a\in \cA$ on which $\alpha$ takes non-zero values, i.e., $\supp{\alpha}:=\{a\in \cA\mid \alpha(a)\neq 0\}$.

Then we denote by $G_\mathbb{C}(\cX)\equiv(\bC^{(\cA)},\cdot)$ the group of \emph{generalized monomials}, whose multiplication $\cdot$ is defined~as
\begin{equation}
\cA^{\alpha}\cdot\cA^{\beta}=\cA^{\alpha+\beta}\,
\end{equation}
here, we have employed the \emph{multi-index notation}
\begin{equation}
\{\cA^\alpha\}_{\alpha\in \mathbb{C}^{(\cA)}}\,
\end{equation}
such that for a given function $\alpha:\cA\rightarrow \bC$ with support
\[
\supp{\alpha}=\{a_1,\dotsc,a_n\}\subset \cA
\]
the corresponding element $\cA^{\alpha}\in G_\mathbb{C}(\cA)$ reads more explicitly
\begin{equation}
\cA^{\alpha}=a_1^{\alpha(a_1)}\dotsc a_n^{\alpha(a_n)}\,
\end{equation}

Moreover, we denote by $\bC[G_{\bC}(\cA)]$ the \emph{$\bC$-algebra} of the group $G_{\bC}(\cA)$.
\end{Definition}

It is instructive to note that for an element $\cA^{\alpha}\in G_\mathbb{C}(\cA)$ with $\supp{\alpha}\subset \bZ$, $\cA^{\alpha}$ is nothing but a Laurent polynomial, while for $\supp{\alpha}\subset \bZ_{\geq 0}$ it is an ordinary polynomial, which motivated the moniker ``generalized monomials'' for generic elements of $G_\mathbb{C}(\cA)$. In other words, a generic element $\cA^{\alpha}\in G_{\bC}(\cA)$ might be thought of as a ``monomial with complex exponents''.

Combining the notion of generalized polynomials (extended to certain formal power series) with the two types of integrals as introduced above, we will now introduce a formal integration operator that will play the central role in our implementation of multi-variate umbral calculus.

\begin{Definition}\label{def:IT}
Let $\cA=\{\lambda\}\uplus\cU\uplus\cV\uplus \cX$ be an alphabet of formal variables, where $\uplus$ denotes the operation of disjoint union, and where $\{\lambda\}$, $\cU$, $\cV$ and $\cX$ are four (disjoint) alphabets of auxiliary formal variables. We will typically employ notations such as $\cX=\{x,y,x_1,x_2,\dotsc\}$, where we make use of the indexed variable notations in case of many variables for convenience. Let furthermore $\cA_\bullet=\cA\setminus\{\lambda\}$.

We define a \emph{formal integration operator} $\tio$ via specifying first its domain $\dom{\tio}$ as
\begin{equation}\label{eq:domainTIO}
\dom{\tio}:=\{S\in \bC[G_\mathbb{C}(\cA_\bullet)][[\lambda]]\mid \mbox{for all } \cA^{\alpha}\in supp(S) : \range{\alpha\vert_{\cU}}\subset\bC\setminus\bZ_{\leq 0}\}\,
\end{equation}
whence elements of $\dom{\tio}$ are formal power series in $\lambda$ with coefficients that are generalized polynomials over the alphabet $\cA_{\bullet}$ (where the extension to formal power series requires a suitable notion of summability, see below). {Here, the notation $\range{\alpha\vert_{\cU}}\subset\bC\setminus\bZ_{\leq 0}$ entails that functions $\alpha:\cA_{\bullet}\rightarrow \bC$ are required to not take zero or negative integer  values when evaluated on elements of $\cU$.} Then for some monomial $\cA^{\alpha}\in \dom{\tio}$, which reads more explicitly (recall that by definition of $\bC^{(\cA)}$, we have that $\alpha(u_i)\neq 0$ and $\alpha(v_j)\neq 0$ for only finitely many indices $i$ and $j$)
\[
\cA^{\alpha}=u_1^{\alpha(u_1)}u_2^{\alpha(u_2)}\dotsc v_1^{\alpha(v_1)}v_2^{\alpha(v_2)}\dotsc\,
\]
the action of $\tio$ on $\cA^{\alpha}$ is defined as
\begin{equation}
\begin{aligned}
\tio\left(\cA^{\alpha}\right)&:=\left(\prod_{u_i \in U(\alpha)}\int_0^{\infty}du_i\; \frac{e^{-u_i}}{u_i}\right)\left[
\left(\prod_{v_j\in V(\alpha)}\frac{1}{2\pi i}\int_{\gamma}dv_j\; e^{v_j}\right)\left[
\cA^{\tilde{\alpha}}\right]\right]\\
U(\alpha)&:=\supp{\alpha}\cap \cU\,,\; V(\alpha):=\supp{\alpha}\cap \cV\\
\tilde{\alpha}(a)&:=\begin{cases}
\alpha(a) &\text{if } a\in \cA\setminus\cV\\
-\alpha(a)\quad &\text{if } a\in \cV\,
\end{cases}
\end{aligned}
\end{equation}

We extend $\tio$ by linearity to finite sums. For infinite sums, this requires an appropriate notion of convergence. A series $\sum_{i\in I}c_i\cA^{\alpha_i}$ will be in $\dom{\tio}$ (the domain of $\tio$) if the family $\Big(c_i\tio(\cA^{\alpha_i})\Big)_{i\in I}$ is \emph{summable} in the target (in the sense of discrete summability or compact convergence for entire functions). For example, in Equation~(\ref{cyl_Bessel1}), the target is the space of entire functions endowed with the topology of compact convergence. In the sequel, the~target will always be a topological ring to ensure continuity of products and sums.
\end{Definition}

Note that our definition of the integration over the formal variables $\cV$ (i.e., integrating the monomial $\cA^{\bar{\alpha}}$ given an input $\cA^{\alpha}$) was designed purely for notational convenience, i.e., such as to avoid having to carry negative exponents in our standard application of e.g., expressing $1/\Gamma(\beta)$ as $\tio(v^{\beta})$ (see also the examples provided in Section~\ref{sec:ucEx} for further illustrations).

The readers may verify that our choice of domain $\dom{\tio}$ as given in~\eqref{eq:domainTIO} precisely ensures that we avoid the poles in the definition of $\Gamma(z)$ in our series transformation calculations, which renders the technique mathematically well-posed.
In all of our practical applications, we will further restrict the formal power series in $\dom{\tio}$ to $\range{\alpha}\cap \cX\subset \bZ_{\geq 0}$, which entails that the \emph{group} of monomials $G_{\bC}(\cA_{\bullet})$ restricts to a \emph{group} for $\cV$, a \emph{semigroup} for $\cX$ and not even a semigroup for $\cU$: for example, consider the admissible monomial $u^{-\frac{1}{2}}$, whose square is not in the domain.

The operator $\tio$ is \emph{multiplicative} on monomials whose underlying alphabets do not overlap, and~it is checked easily that it is not so, in general, for monomials and sums of monomials which have variables in common.  To ensure mathematical consistency, we provide the reader below with a precise statement on multiplicativity.
\begin{Lemma}\label{lem:factorization}
We suppose that
\begin{enumerate}
\item the alphabet $\cA$ is partitioned into two disjoint subalphabets
\[
\cA=\cA_1\uplus \cA_2
\]
\item we are given two series
\[
S_1=\sum_{i\in I_1}c_{\alpha_i}\cA_1^{\alpha_i}\; \text{and}\; S_2=\sum_{j\in I_2}c_{\alpha_j}\cA_2^{\alpha_j}
\]
such that $S_1,S_2\in dom(\tio{})$.
\end{enumerate}

Then $\tio(S_1S_2)$ exists and
\[
\tio(S_1S_2)=\tio(S_1)\tio(S_2)
\]
\end{Lemma}
\begin{proof}
In the sequence of steps below,
\begin{align*}
\tio(S_1)\tio(S_2)&\stackrel{(1)}{=}\sum_{i\in I_1}c_{\alpha_i}\tio(\cA_1^{\alpha_i})\sum_{j\in I_2}c_{\alpha_j}\tio(\cA_2^{\alpha_j})\stackrel{(2)}{=}
\sum_{(i,j)\in I_1\times I_2}c_{\alpha_i}c_{\alpha_j}\tio(\cA_1^{\alpha_i})\tio(\cA_2^{\alpha_j})\\
&\stackrel{(3)}{=}
\sum_{(i,j)\in I_1\times I_2}c_{\alpha_i}c_{\alpha_j}\tio(\cA_1^{\alpha_i}\cA_2^{\alpha_j})
\stackrel{(4)}{=}
\tio\Big(\sum_{(i,j)\in I_1\times I_2}c_{\alpha_i}c_{\alpha_j}\cA_1^{\alpha_i}\cA_2^{\alpha_j}\Big)\stackrel{(5)}{=}\tio(S_1S_2)\,
\end{align*}%
step $(1)$ follows because $S_k,\ k=1,2$ are in $\dom{\tio{}}$, step $(2)$ because the target is a topological ring, step $(3)$ because the alphabets are disjoint, step $(4)$ because the series is summable, and finally step $(5)$ via rewriting of the argument (which is summable for the discrete topology).
\end{proof}

\subsection*{Examples and Relation to Traditional Umbral Calculus}
\label{sec:ucEx}

Consider first as an elementary example the formulae (for $\alpha\in \bC\setminus\bZ_{\leq0}$ and $\beta\in \bC$)
\begin{equation}
\tio\left(u^{\alpha}\right)=\Gamma(\alpha)\,,\quad \tio\left(
v^{\beta}
\right)=\frac{1}{\Gamma(\beta)}\,
\end{equation}

Compared to the umbral calculus literature, we find the analogous equations (see e.g.,~\cite{babusci2013symbolic,babusci2014spherical,dattoli2017operational})
\begin{equation}
\hat{d}^{\gamma}\psi_0:=\Gamma(\gamma+1)\,,\quad \hat{c}^{\delta}\varphi_0:=\frac{1}{\Gamma(\delta+1)}\,
\end{equation}
here, $\psi_0$ and $\varphi_0$ are the ``umbral vacua'', and we have the parameters $\gamma\in \bC\setminus\bZ_{\leq -1}$ and $\delta\in \bC$. The~``umbrae'' $\hat{c}$ and $\hat{d}$ are usually further characterized by formulae such as
\begin{equation}
\hat{c}^{\alpha}\hat{c}^{\beta}\varphi_0=\hat{c}^{\alpha+\beta}\varphi_0\,,\quad \hat{d}^{\gamma}\hat{d}^{\delta}\psi_0=\hat{d}^{\gamma+\delta}\psi_0\,
\end{equation}

However, it is important to note that $\alpha$, $\beta$ and $\alpha+\beta$ must all be elements of $\bC\setminus\bZ_{\leq -1}$, which~renders claims that $\hat{c}$ satisfies some group-like properties as typical in the umbral calculus parlance somewhat questionable. Moreover, typically the precise typing of the ``vacua'' is left implicit, thereby adding to the difficulties in validating umbral techniques beyond individual application examples. The nature of the ``vacua'' becomes even more implicit to characterize upon considering multi-variate extensions of the umbral methods, the analogues of which in our own formulation will find frequent application in this paper. Here, we thus prefer to work instead solely based on the definition of the operation $\tio$, taking care in each individual example that $\tio$ is applied to a well-formed expression in its domain $D(\lambda;\cU,\cV;\cX)$.

To provide a few first application examples for our integral operator approach seen to manipulate formal power series, note first that by definition (for $\alpha,\beta\in \bC\setminus\bZ_{\leq 0}$ and $n\in \bZ_{\geq 0}$)
\begin{equation}\label{eq:poch}
\tio\left(u^{\alpha+n}v^{\alpha}\right)=\frac{\Gamma(\alpha+n)}{\Gamma(\alpha)}=(\alpha)_n\,,\quad
\tio\left(u^{\beta}v^{\beta+n}\right)=\frac{\Gamma(\beta)}{\Gamma(\beta+n)}=\frac{1}{(\beta)_n}\,
\end{equation}
here, we employ the notation $(x)_y:=\Gamma(x+y)/\Gamma(x)$ for the \emph{Pochhammer symbol}, as customary e.g., in the literature on hypergeometric functions. The relations provided in~\eqref{eq:poch} induce an interesting alternative formula for the generalized hypergeometric functions, with parameters $\alpha_i,\beta_j\in \bC\setminus\bZ_{\leq 0}$, and where we consider the variable $z$ as \emph{formal} in order to avoid issues of convergence (see e.g.,~\cite{DLMF}, §16.2, according to which in particular all generalized hypergeometric series with $p>q+1$ are~divergent):
\begin{equation}
\pFq{p}{q}{\vec{\alpha}}{\vec{\beta}}{z}\equiv\pFq{p}{q}{\seq{\alpha_i}{1\leq i\leq p}}{\seq{\beta_j}{1\leq j\leq q}}{z}:=\sum_{n\geq 0}\frac{z^n}{n!}\frac{(\alpha_1)_n\cdots (\alpha_p)_n}{(\beta_1)_n\cdots (\beta_q)_n}\
\end{equation}
namely
\begin{equation}\label{eq:pFqSIT}
\pFq{p}{q}{\vec{\alpha}}{\vec{\beta}}{z}=
\tio\left(
\left(\prod_{i=1}^p (u_i v_i)^{\alpha_i}\right)
\left(\prod_{j=1}^q (u_{j+p} v_{j+p})^{\beta_j}\right)\;e^{z u_1\cdots u_p v_{p+1}\cdots v_{p+q}}
\right)\,
\end{equation}

To demonstrate the utility of the integral transform approach as a method to study special functions, we next present an adaptation of a result obtained with umbral methods in~\cite{babusci2014spherical}. Consider~first the following computation:
\begin{subequations}
\begin{align}
\tio(G_{v}(x))&:=\tio\left(ve^{-v\left(\frac{x}{2}\right)^2}\right)\\
&=\tio\left(
\sum_{r\geq 0}\frac{(-1)^r}{r!}\left(\frac{x}{2}\right)^{2r}v^{r+1}
\right)\\
&=\sum_{r\geq 0}\frac{(-1)^r}{(r!)^2}\left(\frac{x}{2}\right)^{2r}=J_0(x)\,.
\end{align}
\end{subequations}
here, $J_0(x)$ denotes the $0$-th order cylindrical Bessel function. We have thus established that the so-called \emph{umbral image} of the Gaussian function $G_{\alpha}(x):=\alpha\exp(-\alpha (x/2)^2)$ (at $\alpha=v$ and integrated using the operation $\tio$) is the function $J_0(x)$. This finding may be further generalized to the following relation involving the $n$-th order cylindrical Bessel functions $J_n(x)$:
\begin{equation}\label{cyl_Bessel1}
\tio\left(v\left(\frac{vx}{2}\right)^n
e^{-v\left(\frac{x}{2}\right)^2}
\right)=\sum_{r=0}^{\infty}\frac{(-1)^r}{(n+r)!r!}\left(\frac{x}{2}\right)^{n+2r}=J_n(x)\,
\end{equation}

As explained in~\cite{babusci2014spherical}, such a restyling has allowed noticeable simplifications, concerning the handling of integrals of Bessel functions and of many other problems associated with the use of the Ramanujan master theorem. For the purposes of the present paper, a multi-variate analogue of the ``umbral image'' concept will permit us to uncover some hidden structures in the study of SJ~polynomials.

Our methods inherit several equivalences from the properties of the gamma and reciprocal gamma functions, which depending on the application may pose an obstacle or a virtue but are in a certain sense unavoidable. Based on the functional relations
\[
\Gamma(z+1)=z\Gamma(z)\qquad (z\in \bC \setminus{\bZ_{\leq 0}})\,
\]
we find the umbral image analogues (for $\alpha,\beta\in \bC\setminus\bZ_{\leq 0}$)
\begin{equation}\label{eq:auxUV}
\begin{aligned}
\tio\left(e^{\frac{\partial}{\partial a}}u^a
\right)\big\vert_{a\to\alpha}&= \tio\left(
u^{\alpha+1}
\right)=\tio\left(
\alpha u^{\alpha}
\right)=\tio\left(\tfrac{\partial}{\partial u}\, u^{\alpha}\right)\\
\tio\left(e^{\frac{\partial}{\partial b}}v^b
\right)\big\vert_{b\to\beta}&= \tio\left(
v^{\beta+1}
\right)=\tio\left(
\frac{v^{\beta}}{\beta}
\right)\,
\end{aligned}
\end{equation}

For convenience, we will from hereon take a notational convention for formal derivatives that permits to rewrite the first line of the above equation in the more concise form
\begin{equation}
e^{\partial_{\beta}}v^{\beta}
\equiv\left(e^{\frac{\partial}{\partial b}}v^b
\right)\big\vert_{b\to\beta}\,
\end{equation}

As an example, for interesting consequences of such identities, consider Euler's beta function~(\cite{DLMF}, §5.12) $B(\alpha,\beta)$
\begin{equation}\label{eq:betaDef}
B(\alpha,\beta)=\frac{\Gamma(\alpha)\Gamma(\beta)}{\Gamma(\alpha+\beta)}=\tio\left(u_1^{\alpha}u_2^{\beta}v^{\alpha+\beta}\right)\,
\end{equation}
here and throughout the remainder of this section, we assume that $\alpha$ and $\beta$ are in the domain of definition of $B(\alpha,\beta)$, whence that $\alpha,\beta\in \bC\setminus\bZ_{\leq0}$ with $(\alpha+\beta)\in \bC\setminus\bZ_{\leq0}$. Owing to its well-known property (obtained by multiplying numerator and denominator by $(\alpha+\beta)$)
\begin{equation}\label{eq:betaprop}
B(\alpha,\beta)=\frac{(\alpha+\beta)\Gamma(\alpha)\Gamma(\beta)}{(\alpha+\beta)\Gamma(\alpha+\beta)}=
\frac{\alpha\Gamma(\alpha)\Gamma(\beta)}{\Gamma(\alpha+\beta+1)}
+\frac{\beta\Gamma(\alpha)\Gamma(\beta)}{\Gamma(\alpha+\beta+1)}
=B(\alpha+1,\beta)+B(\alpha,\beta+1)\,,
\end{equation}
we may identify $B(\alpha,\beta)$ as eigenfunctions of eigenvalue $1$ of the differential operator $e^{\partial_\alpha}+e^{\partial_{\beta}}$,
\begin{subequations}\label{eq:betaProp2}
\begin{align}
B(\alpha,\beta)&=\left(e^{\partial_\alpha}+e^{\partial_{\beta}}\right)B(\alpha,\beta)\\
\Leftrightarrow\quad
\tio\left(u_1^{\alpha}u_2^{\beta}v^{\alpha+\beta}\right)&=
\tio\left(\left(e^{\partial_\alpha}+e^{\partial_{\beta}}\right)u_1^{\alpha}u_2^{\beta}v^{\alpha+\beta}\right)
=\tio\left(u_1^{\alpha}u_2^{\beta}v^{\alpha+\beta+1}(u_1+u_2)\right) \,
\end{align}
\end{subequations}

We find in fact an infinite tower of equivalences by noting that for any non-negative integer $n\in \bZ_{\geq 0}$,
\begin{subequations}
\begin{align}
B(\alpha,\beta)&=\left(e^{\partial_\alpha}+e^{\partial_{\beta}}\right)^nB(\alpha,\beta)\\
\Leftrightarrow\quad
\tio\left(u_1^{\alpha}u_2^{\beta}v^{\alpha+\beta}\right)&=
\tio\left(\left(e^{\partial_\alpha}+e^{\partial_{\beta}}\right)^nu_1^{\alpha}u_2^{\beta}v^{\alpha+\beta}\right)
=\tio\left(u_1^{\alpha}u_2^{\beta}v^{\alpha+\beta+n}(u_1+u_2)^n\right) \,
\end{align}
\end{subequations}

Moreover, by rephrasing~\eqref{eq:betaprop} in the form
\begin{equation}
B(\alpha,\beta+1)=B(\alpha,\beta)-B(\alpha+1,\beta)=\left(1-e^{\partial_{\alpha}}\right)B(\alpha,\beta)
\end{equation}
iterating yields the family of identities (for $n\in \bZ_{\geq 0}$)
\begin{subequations}
\begin{align}
B(\alpha,\beta+n)&=\left(1-e^{\partial_{\alpha}}\right)^nB(\alpha,\beta)\\
\Leftrightarrow\quad
\tio\left(u_1^{\alpha}u_2^{\beta+n}v^{\alpha+\beta+n}\right)&=
\tio\left(\left(1-e^{\partial_{\alpha}}\right)^n u_1^{\alpha}u_2^{\beta}v^{\alpha+\beta}\right)
=
\tio\left(u_1^{\alpha}u_2^{\beta}v^{\alpha+\beta}\left(1-u_1v\right)^n\right)\,\label{eq:appDaux}
\end{align}
\end{subequations}

The last Equation~\eqref{eq:appDaux} will prove to be of particular importance for the forthcoming discussion. It will be exploited to settle out the proof of important  identities, whose technicalities are detailed e.g., in Appendix~\ref{app:AB}.

We conclude this section with a particularly interesting use of the auxiliary relations~\eqref{eq:auxUV}, regarding~explicit expressions for the derivative of generalized hypergeometric functions ${}_pF_q(\vec{\alpha},\vec{\beta};z)$ (which we will need in our calculations of shifts of lacunary generating functions in Section~\ref{sec:LGSJ}). Acting~with a derivative by $z$ on~\eqref{eq:pFqSIT} and using the auxiliary relations~\eqref{eq:auxUV} repeatedly leads to:
\begin{equation}
\begin{aligned}
\tfrac{\partial}{\partial z}\pFq{p}{q}{\vec{\alpha}}{\vec{\beta}}{z}&=
\tio\left(
\left(\prod_{i=1}^p (u_i v_i)^{\alpha_i}\right)
\left(\prod_{j=1}^q (u_{j+p} v_{j+p})^{\beta_j}\right)\;u_1\cdots u_p v_{p+1}\cdots v_{p+q}e^{z u_1\cdots u_p v_{p+1}\cdots v_{p+q}}
\right)\\
&=\left(\frac{\prod_{i=1}^p \alpha_i}{\prod_{j=1}^q\beta_j}\right)\tio\left(
\left(\prod_{i=1}^p (u_i v_i)^{\alpha_i+1}\right)
\left(\prod_{j=1}^q (u_{j+p} v_{j+p})^{\beta_j+1}\right)\;e^{z u_1\cdots u_p v_{p+1}\cdots v_{p+q}}
\right)\,
\end{aligned}
\end{equation}

We thus find the expression
\begin{equation}
\tfrac{\partial}{\partial z}\pFq{p}{q}{\vec{\alpha}}{\vec{\beta}}{z}=\left(\frac{\prod_{i=1}^p \alpha_i}{\prod_{j=1}^q\beta_j}\right)\pFq{p}{q}{\vec{\alpha+1}}{\vec{\beta+1}}{z}\,.
\end{equation}

Proceeding analogously for higher $z$-derivatives then reproduces the following closed-form expression (see Chapter~1.30.1.1 in~\cite{brychkov2008handbook}), for $n\in \bZ_{>0}$:
\begin{equation}
\tfrac{\partial^n}{\partial z^n}\pFq{p}{q}{\vec{\alpha}}{\vec{\beta}}{z}=\left(\frac{\prod_{i=1}^p (\alpha_i)_n}{\prod_{j=1}^q(\beta_j)_n}\right)\pFq{p}{q}{\vec{\alpha+n}}{\vec{\beta+n}}{z}\,
\end{equation}

\section{The Sobolev-Jacobi Polynomials}
\label{sec:SJP}

We introduce (following Kwon and Littlejohn~\cite{kwon1994characterizations,kwon1996new,kwon1998sobolev}) the \emph{SJ polynomials} $\tilde{P}^{(-1,-\beta)}_n(x)$ as the polynomial eigenfunctions of the Jacobi-type differential operators $\hat{D}^{(-1,\beta)}_{Jac}$ (with $\beta\in \bR_{\geq -1}$):
\begin{equation}
\label{SJ:def}
\begin{aligned}
\hat{D}^{(-1,\beta)}_{Jac}\tilde{P}^{(-1,\beta)}_n(x)
&=
-n(n+\beta)\tilde{P}^{(-1,\beta)}_n(x)\\
\hat{D}^{(-1,\beta)}_{Jac}&=(1-\hat{x}^2)\partial_x^2
+(\beta+1)\big(1-\hat{x}\big)\partial_x\,
\end{aligned}
\end{equation}
here, we have made use of the notations $\hat{x}$ and $\partial_x$ for the following linear operators:
\begin{equation}
\hat{x}f(x):=xf(x)\,,\quad \partial_x f(x)=\tfrac{\partial}{\partial x}f(x)\,
\end{equation}

It will prove useful in our later considerations that $\hat{x}$ and $\partial_x$ form a representation of the \emph{Heisenberg-Weyl algebra} (i.e., the \emph{Bargmann-Fock representation}):
\begin{equation}\label{eq:HWrep}
[\partial_x,\hat{x}]=\mathbbm{1}\,
\end{equation}

For clarity, let us briefly comment on the relationship of the SJ polynomials to the ordinary Jacobi polynomials $P^{(\alpha,\beta)}_n(x)$. The Jacobi-type differential operator(cf.\ the standard reference book~\cite{olver2010nist}, chapter 18, p.~445, Table~18.8.1)
\begin{equation}
\label{eq:DEjac}
\begin{aligned}
\hat{D}^{(\alpha,\beta)}_{Jac}&:=(1-x^2)\partial_x^2
+q^{(\alpha,\beta)}(x)\partial_x\\
q^{(\alpha,\beta)}(x)&=(\beta-\alpha-(\alpha+\beta+2)x)
\end{aligned}
\end{equation}
with \emph{real parameters}
\begin{equation}
\label{eq:Jparam}
\alpha,\beta>-1
\end{equation}
is known to possess a \emph{system of orthogonal polynomials} $\{P^{(\alpha,\beta)}_n(z)\}_{n\geq 0}$, the so-called \emph{classical Jacobi polynomials}, as its complete basis of eigenfunctions (see~\eqref{eq:Jacorth} for the associated inner product). More~explicitly (cf.\ e.g., Equation~(18.5.8) in \cite{olver2010nist}),
\begin{equation}\label{eq:JacExpl}
P^{(\alpha,\beta)}_n(x)=
\sum_{\ell=0}^n\binom{n+\alpha}{\ell}\binom{n+\beta}{n-\ell}
\left(\frac{x-1}{2}\right)^{n-\ell}\left(\frac{x+1}{2}\right)^{\ell}\,
\end{equation}
satisfying the eigenequation
\begin{equation}\label{eq:JacEigen}
\hat{D}^{(\alpha,\beta)}_{Jac}P_n^{(\alpha,\beta)}(x)=-n(n+\alpha +\beta+1)P_n^{(\alpha,\beta)}(x)\,
\end{equation}

The orthogonality property is found by defining for each admissible choice of parameters \mbox{$\alpha,\beta>-1$} an \emph{inner product} $\Phi_{\alpha,\beta}$ on the space of polynomials $\bR[x]$,
\begin{subequations}\label{eq:Jacorth}
\begin{align}
\label{eq:innerJclass}
\Phi_{\alpha,\beta}(p,q)&:=\int_{-1}^{+1}dx\; w_{\alpha,\beta}(x)p(x)q(x)\\
w_{\alpha,\beta}(x)&:=(1-x)^{\alpha}(1+x)^{\beta}\,,\; p,q\in \bR[x]\,
\end{align}
\end{subequations}

It is one of the classical results of the theory of orthogonal polynomials that for all $n\geq 0$
\begin{equation}
P_{n}^{(\alpha,\beta)}(x)\in L^2_{\alpha,\beta}([-1,1])\,
\end{equation}
here, $L^2_{\alpha,\beta}([-1,1])$ denotes the space of functions on the interval $[-1,1]$ which are \emph{square-integrable} with respect to the integral against the weight function $w_{\alpha,\beta}(z)$. Moreover, orthogonality manifests itself as
\begin{equation}
\Phi_{\alpha,\beta}\left(P_m^{(\alpha,\beta)},P_n^{(\alpha,\beta)}\right)=\delta_{m,n}\phi_{\alpha,\beta}(n)\,
\end{equation}
where $\phi_{\alpha,\beta}(n)\in \bR_{>0}$ are some (non-zero) real numbers.

However, if one wishes to study a Jacobi-type differential operator with $\alpha=-1$ and either $\beta>-1$ or $\beta=-1$, the family of polynomials $\{P_n^{(\alpha,\beta)}(x)\}_{n\in \bZ_{\geq0}}$ ceases to constitute a complete system of orthogonal polynomials: for both cases, the polynomial $P_0^{(-1,\beta)}(x)=1$ is not integrable, and moreover in the case $\alpha=-1$ and $\beta=-1$ one finds that $P_1^{(-1,-1)}(x)=0$. Referring to~\cite{bdp2017} for a detailed discussion, the resolution of this problem proposed by Kwon and Littlejohn~\cite{kwon1994characterizations,kwon1996new,kwon1998sobolev,Bruder_2012} consists in redefining the notion of orthogonality: one no longer requires orthogonality of the polynomials with respect to the inner product $\Phi_{\alpha,\beta}$, but instead defines a Sobolev-type inner product. We quote from~\cite{kwon1998sobolev} the relevant material (cf.\ Propositions 4.2 and 4.3 of loc cit):
\begin{Proposition}\label{prop:KL}
Let the \emph{Sobolev inner products}   $\Phi^{(-1,-1)}_{A,B}$ and $\Phi_C^{(-1,\beta)}$ (for $\beta>-1$) be defined as follows (for~$p,q\in \bR[x]$ some polynomials with real coefficients):
\begin{equation}
\begin{aligned}
\Phi^{(-1,-1)}_{A,B}(p,q)&:=A p(1)q(1)+B p(-1)q(-1) +\int_{-1}^{+1}dx\; p'(x)q'(x)\\
\Phi^{(-1,\beta)}_C(p,q)&:= C p(1)q(1)+\int_{-1}^{+1}dx\; (x+1)^{\beta+1}p'(x)q'(x)\qquad (\beta>-1)\,
\end{aligned}
\end{equation}
here, $p'(x)$ and $q'(x)$ denote the first derivatives of the polynomials, and $A,B,C\in \bR$ are parameters. For~the resulting inner products to be \emph{positive definite}, the parameters $A,B,C\in \bR$ must satisfy the following conditions: for $\Phi^{(-1,-1)}_{A,B}$, $A$, and $B$ must verify
\begin{equation}
A+B>0\,,\; A(\gamma+1)^2+B(\gamma-1)^2+2\neq 0\,,\; A(\gamma+1)+B(\gamma-1)=0\,,\; \gamma:=(B-A)/(A+B)\,
\end{equation}
while for $\Phi_C^{(-1,\beta)}$ the parameter $C$ must verify
\begin{equation}
C>0\,
\end{equation}

Then the \emph{SJ orthogonal polynomials} $\tilde{P}^{(\alpha,\beta)}_n(x)$ are defined for $\alpha=-1,~\beta=-1$ as
\begin{equation}\label{eq:SJmm}
\begin{aligned}
&\tilde{P}^{(-1,-1)}_0(x):=1\\
&\tilde{P}^{(-1,-1)}_1(x):=x+\gamma\\
&\tilde{P}^{(-1,-1)}_{n\geq 2}(x):=\binom{2n-2}{n}^{-1}\sum_{k=1}^{n-1}\binom{n-1}{k}\binom{n-1}{n-k}(x-1)^{n-k}(x+1)^k\,
\end{aligned}
\end{equation}
where $\gamma=(B-A)/(A+B)$, while for the parameters $\alpha=-1,~\beta>-1$ one defines
\begin{equation}
\begin{aligned}\label{eq:SJbeta}
&\tilde{P}_0^{(-1,\beta)}(x):=1\\
&\tilde{P}_1^{(-1,\beta)}(x):=x-1\\
&\tilde{P}_{n\geq 2}^{(-1,\beta)}(x):=\binom{2n+\beta-1}{n}^{-1}\sum_{k=0}^n\binom{n-1}{k}\binom{n+\beta}{n-k}(x-1)^{n-k}(x+1)^k\,
\end{aligned}
\end{equation}

The polynomials $\tilde{P}^{(\alpha,\beta)}_n(x)$ form a complete orthogonal system of polynomial eigenfunctions of the Jacobi differential operator at parameters in the aforementioned parameter ranges.
\end{Proposition}

The readers may have noticed that indeed the SJ polynomials $\tilde{P}^{(\alpha,\beta)}_n(x)$ as defined above coincide for $n\geq 2$ with the \emph{monic versions of the classical Jacobi polynomials} (i.e., a different normalization choice where in each polynomial of degree $n$ the coefficient of $x^n$ is normalized to be $1$).

There exists an alternative approach to the definition of the SJ polynomials due to Gurappa and Panigrahi~\cite{gurappa1996new,gurappa1999equivalence,gurappa2001novel,gurappa2002linear,Shreecharan_2004,gurappa2004polynomial}, which will prove quintessential to our present paper. Quite remarkably, their technique of solving rather general differential equations by providing a constructive algorithm to determine polynomial eigenfunctions of the respective differential operators leads to a \emph{unified construction} of classical \emph{and} SJ monic orthogonal polynomials (albeit the authors do not appear to highlight this feature explicitly in their articles). Since there unfortunately appears to occur a systematic typographic error in loc.\ cit.\ in the equations pertaining to Jacobi polynomials, we will now present a careful re-derivation of the relevant results.

Specializing the general technique of Gurappa and Panigrahi to the Jacobi-type differential Equations~\eqref{SJ:def}and~\eqref{eq:DEjac}, consider an eigenequation of the form~\eqref{eq:JacEigen}, where we however from here on permit the parameter ranges $\alpha,\beta\in \bR_{\geq -1}$ (i.e., explicitly including the special cases $\alpha=-1$ and $\beta=-1$ or $\beta>-1$ discussed above). We will temporarily employ the notation $\tilde{P}^{(\alpha,\beta)}_n(x)$ for this extended parameter range (with $\tilde{P}^{(\alpha,\beta)}_n(x):=P^{(\alpha,\beta)}_n(x)$ if $\alpha>-1$ and $\beta>-1$). The first step in their technique consists in a certain splitting of the linear operators in the eigenequation~\eqref{eq:JacEigen} into a ``diagonal'' part $F^{(\alpha,\beta)}_n(\hat{D}_x)$ and an ``non-diagonal'' part $N(\hat{x},\partial_x)$ (where we chose to use the notation $N(\dotsc)$ for the ``non-diagonal'' parts rather than the notation $P(\dotsc)$ used in loc cit.\ in order to avoid any potential confusion of the operator $N(\dotsc)$ with polynomials):
\begin{equation}\label{eq:opSplit}
\begin{aligned}
0&=\left(\hat{D}^{(\alpha,\beta)}_{Jac}+n(n+\alpha+\beta+1)\right)\tilde{P}^{(\alpha,\beta)}_n(x)\\
&=(F^{(\alpha,\beta)}_n(\hat{D}_x)+N(\hat{x},\partial_x))\tilde{P}^{(-1,\beta)}_n(x)\,
\end{aligned}
\end{equation}
Here, $\hat{D}_x:=\hat{x}\partial_x$ denotes the \emph{Euler operator}, whence the \emph{diagonal operator} with action on monomials
\begin{equation}
\hat{D}_x x^n=n x^n\,
\end{equation}

Using the canonical commutation relation~\eqref{eq:HWrep} to compute the auxiliary formula
\begin{equation}\label{eq:EulerSq}
\hat{D}_x^2=\hat{x}^2\partial_x^2+\hat{D}_x\,
\end{equation}
the concrete expressions for the operators $F^{(\alpha,\beta)}_n(\hat{D}_x)$ and $N(\hat{x},\partial_x)$ in~\eqref{eq:opSplit} read (see Appendix~\ref{app:SJdef} for the details of the derivation; compare~\cite{gurappa2004polynomial}, which however contains a slight error in the respective formula for $F^{(\alpha,\beta)}_n(\hat{D}_x)$):
\begin{equation}
\begin{aligned}
F^{(\alpha,\beta)}_n(\hat{D}_x)&:=
-(\hat{D}_x-n)(\hat{D}_x+n+\alpha+\beta+1)
\\
N(\hat{x},\partial_x)&=\partial_x^2+(\beta-\alpha)\partial_x\,
\end{aligned}
\end{equation}

The key idea of the technique of Gurappa and Panigrahi consists then in recognizing the following relationship:
\begin{equation}\label{eq:auxRel}
F^{(\alpha,\beta)}_n(\hat{D}_x)x^n=0\,
\end{equation}

Moreover, $F^{(\alpha,\beta)}_n(\hat{D}_x)$ acts as a \emph{diagonal} operator on any monomial, which allows to perform the following computations (with shorthand notations $\hat{F}_n\equiv F^{(\alpha,\beta)}_n(\hat{D}_x)$, considered temporarily
as a symbol, and $\hat{N}\equiv N(\hat{x},\partial_x)$):
\begin{equation}
\begin{aligned}
0&=(\hat{F}_n+\hat{N})\tilde{P}^{(-1,\beta)}_n(x)=\hat{F}_n(1+\tfrac{1}{\hat{F}_n}\hat{N})\tilde{P}^{(-1,\beta)}_n(x)\,
\end{aligned}
\end{equation}
where it is important to note that $\tfrac{1}{\hat{F}_n}$ and $\hat{N}$ \emph{do not commute}, whence the ordering in the last equation is important. The final step of the technique then consists in recognizing the following expression for the eigenfunctions:
\begin{subequations}\label{eq:SJexpl}
\begin{align}
&\tilde{P}^{(\alpha,\beta)}_n(x)=c_n\frac{1}{1+\tfrac{1}{\hat{F}_n}\hat{N}}x^n=c_n\sum_{m\geq0}(-1)^m\left[\tfrac{1}{\hat{F}_n}\hat{N}\right]^m x^n\\
&\quad=c_n\sum_{m\geq0}\left[\tfrac{1}{(\hat{D}_x-n)(\hat{D}_x+n+\alpha+\beta +1)}\left(\partial_x^2+(\beta-\alpha)\partial_x\right)\right]^m x^n\,.
\end{align}%
\end{subequations}
here, {the non-zero real parameter} $c_n\in \bR\setminus\{0\}$ is a normalization constant. The proof that the functions $\tilde{P}^{(\alpha,\beta)}_n(x)$ as defined in~\eqref{eq:SJexpl} indeed solve the eigenequation is straightforward (cf.\ e.g.,~\cite{gurappa2002linear}):
\begin{equation*}
\begin{aligned}
(\hat{F}_n+\hat{N})\tilde{P}^{(\alpha,\beta)}_n(x)
&=\hat{F}_n(1+\tfrac{1}{\hat{F}_n}\hat{N})
c_n\frac{1}{\left(1+\tfrac{1}{\hat{F}_n}\hat{N}\right)}x^n\\
&=c_n\hat{F}_n(1+\tfrac{1}{\hat{F}_n}\hat{N})
\sum_{m\geq0}(-1)^m\left[\tfrac{1}{\hat{F}_n}\hat{N}\right]^m x^n\\
&=c_n\hat{F}_nx^n\overset{\eqref{eq:auxRel}}{=}0\,
\end{aligned}
\end{equation*}

Since in applications we will be interested in studying the \emph{monic} SJ polynomials (i.e., with leading coefficient equal to $1$), we will set $c_n=1$ for all $n\in \bZ_{\geq 0}$, and summarize the findings of this section with the following proposition:
\begin{Proposition}\label{prop:SJ1}
The \emph{(monic) SJ polynomials} $\tilde{P}^{(-1,\beta)}_n(x)$ (with $\beta\in \bR_{\geq -1}$ and $n\in \bZ_{\geq 0}$) and the \emph{classical monic Jacobi polynomials} $\tilde{P}^{(\alpha,\beta)}_n(x)$ (for $\alpha,\beta\in \bR_{>-1}$), defined as
\begin{equation}\label{eq:SJGP}
\begin{aligned}
\tilde{P}^{(\alpha,\beta)}_n(x)&=\sum_{m\geq0}\left[\tfrac{1}{(\hat{D}_x-n)(\hat{D}_x+n+\alpha+\beta +1)}\left(\partial_x^2+(\beta-\alpha)\partial_x\right)\right]^m x^n\,
\end{aligned}
\end{equation}
are \emph{polynomial eigenfunctions} of the Jacobi-type differential operator $\hat{D}^{(\alpha,\beta)}_n$ with eigenvalues
\begin{equation}
\lambda_n^{(\alpha,\beta)}=-n(n+\alpha+\beta+1)\,
\end{equation}
The  degeneracies of the denominator in~\eqref{eq:SJGP} are counteracted by the derivatives, so the statement can be made rigorous by completing this operator with identity on its kernel: for $\alpha+\beta+1\neq 0$ we have that
\[
ker((\hat{D}_x-n)(\hat{D}_x+n+\alpha+\beta+1))=\bC\, x^n\,
\]
and for $\alpha+\beta+1=0$
\[
ker((\hat{D}_x-n)(\hat{D}_x+n+\alpha+\beta+1))=\bC\, x^n\uplus \bC \,1\,
\]
Completing the operator
by the identity map on its kernel renders $(\hat{D}_x-n)(\hat{D}_x+n+\alpha+\beta+1)$
invertible. The readers may verify easily that $Id$ can be replaced by any invertible endomorphism $N\to N$ without modifying the final result of Equation~\eqref{eq:SJGP}.
\end{Proposition}

The detailed analysis of the two cases (i) $\alpha=\beta=-1$, and (ii) $\alpha=-1$ and $\beta>-1$ will reveal that the corresponding families of SJ polynomials have in fact a rather distinct structure. Accordingly, they~will be treated separately in the following.

\section{SJ Polynomials for $\mathbold{\alpha=-1}$ and $\mathbold{\beta>-1}$}
\label{sec:SJalt}

Compared to the SJ polynomials with parameters $\alpha=\beta=-1$ as discussed in the next section, the~case $\alpha=-1$ and $\beta>-1$ is somewhat more intricate from the viewpoint of the technique of Gurappa Panigrahi. The origin of this difficulty (compare also the results of Gurappa, \mbox{Panigrahi et al.}\ presented in~\cite{gurappa2007applications}) resides in the fact that the SJ polynomials for the case $\alpha=-1$ and $\beta>-1$ may not be expressed in the form of the exponential of a differential operator acting on a monomial (in contrast to the case $\alpha=\beta=-1$, see Proposition~\ref{prop:SJexp}). On the other hand, as evident from the explicit formula as given in~\eqref{eq:SJbeta} (see also~\cite{bdp2017} and references given therein for further details) for these polynomials, they~coincide precisely with the \emph{ordinary} (monic) Jacobi polynomials (while of course their orthogonality property is a different one).~Therefore, while we postpone a more detailed analysis of the connection coefficients for these polynomials to future work, we~briefly present a derivation of the EGF for this case, which is based on~\cite{dattoli2000generalized,dattoli2010generalized}.

Note first that the results presented in~~\eqref{eq:SJbeta} entail that the formula for $\tilde{P}^{(-1,\beta)}_1(x)$ may alternatively be obtained by setting $n=1$ in the formulae for $\tilde{P}^{(-1,\beta)}_{n\geq2}(x)$. Making use of the integral transform techniques introduced in Section~\ref{sec:ST}, we may rewrite the formula for $\tilde{P}^{(-1,\beta)}_{n\geq1}(x)$ as follows:
\begin{equation}
\begin{aligned}
\tilde{P}^{(-1,\beta)}_{n\geq1}(x)&=
\binom{2n+\beta-1}{n}^{-1} \Gamma(n+\beta+1)
\sum_{k=0}^{n-1}\binom{n-1}{k}\frac{(x-1)^{n-k}(x+1)^k}{\Gamma(n-k+1)\Gamma(\beta+k+1)}\\
&=\binom{2n+\beta-1}{n}^{-1} \Gamma(n+\beta+1)(x-1)\;\cdot\\
&\quad
\tio\left(
\sum_{k=0}^{n-1}\binom{n-1}{k}(x-1)^{n-1-k}(x+1)^k v_1^{n-k+1}v_2^{\beta+k+1}
\right)\\
&=\binom{2n+\beta-1}{n}^{-1} \Gamma(n+\beta+1)(x-1)
\tio\left(
v_1^2v_2^{\beta+1}\left(v_1(x-1)+v_2(x+1)\right)^{n-1}
\right)\,
\end{aligned}
\end{equation}

To formulate generating functions for these polynomials, it will prove convenient to work with a rescaled variant $P^{(-1,\beta)}_{n}(x)$ of the SJ polynomials $\tilde{P}^{(-1,\beta)}_{n}(x)$, defined as follows:
\begin{equation}
P^{(-1,\beta)}_{0}(x):=\tilde{P}^{(-1,\beta)}_{0}(x)=1\,,\quad
P^{(-1,\beta)}_{n\geq1}(x):=\binom{2n+\beta-1}{n} \tfrac{1}{\Gamma(n+\beta+1)}\tilde{P}^{(-1,\beta)}_{n\geq1}(x)\,
\end{equation}
with these preparations, it is possible to compute the \emph{$1$-shifted EGF} $\cG_{P_{1,1}}^{(-1,\beta)}(\lambda;x)$ of the rescaled SJ polynomials as follows:
\begin{subequations}
\begin{align}
\cG_{P_{1,1}}^{(-1,\beta)}(\lambda;x)&:=\sum_{n\geq 0}\frac{\lambda^n}{n!} P^{(-1,\beta)}_{n+1}(x)\\
&=(x-1)\tio\left(
v_1^2v_2^{\beta+1}\sum_{n\geq 0}\frac{\lambda^n}{n!}\left(v_1(x-1)+v_2(x+1)\right)^{n}
\right)\\
&=(x-1)\tio\left(
v_1^2v_2^{\beta+1}e^{\lambda \left(v_1(x-1)+v_2(x+1)\right)}
\right)\\
&\overset{*}{=}
(x-1)\left(\tio\left(
v_1^2e^{\lambda v_1(x-1)}
\right)\right)\left(\tio\left(
v_2^{\beta+1}e^{\lambda v_2(x+1)}
\right)\right)\\
&=(x-1) C_1(-\lambda(x-1))C_{\beta}(-\lambda(x+1))\,
\end{align}
\end{subequations}
here, in the step marked $(*)$, we have made use of Lemma~\ref{lem:factorization}, i.e., of the fact that the integrations over $v_1$ and $v_2$ are independent of one another. Moreover, in the last step, we defined the auxiliary functions (for {a complex parameter $\alpha$ that is non-zero and not equal to a negative integer, $\alpha\in \bC\setminus \bZ_{<0}$, and~with} formal variable $z$) known as \emph{Tricomi-Bessel functions}~\cite{srivastavatreatise} (which are related to the modified Bessel functions of the first kind $I_{\alpha}(x)$ as shown below)
\begin{equation}
C_{\alpha}(-z):=\sum_{r\geq0}\frac{z^r}{r! \Gamma(r+\alpha+1)}=z^{-\alpha/2}I_{\alpha}(2\sqrt{z})=\tio\left(
v^{\alpha+1}e^{vz}
\right)\,
\end{equation}

Finally, proceeding as above to define a non-shifted EGF for the polynomials $P^{(-1,\beta)}_{n}(x)$, we may derive the following (formal) equation:
\begin{subequations}
\begin{align}
\cG_{P_{1,0}}^{(-1,\beta)}(\lambda;x)&:=\sum_{n\geq 0}\frac{\lambda^n}{n!} P^{(-1,\beta)}_{n}(x)\\
&=1+\lambda(x-1)\tio\left(
v C_1(-\lambda uv(x-1))C_{\beta}(-\lambda uv(x+1))
\right)\,
\end{align}
\end{subequations}

\section{SJ Polynomials for $\mathbold{\alpha=\beta=-1}$}
\label{sec:SJmain}

Before proceeding to a detailed study of the SJ polynomials for the parameters $\alpha=\beta=-1$ in the remainder of this paper, we will first introduce several useful technical tools.

\subsection{Euler Operators and Normal-Ordering Type Techniques}
\label{sec:ENO}

Denoting by $\hat{D}_x:=\hat{x}\partial_x$ the Euler operator as before, it is straightforward to derive the following auxiliary normal-ordering type identities:
\begin{Lemma}\label{lem:ENO}
For any entire function $f\equiv f(\hat{D}_x)$ in the Euler operator $\hat{D}_x$, and for any non-negative integers $p,q\in \bZ_{\geq0}$, it holds that
\begin{subequations}\label{eq:ENO1}
\begin{align}
f(\hat{D}_x)\hat{x}^p\partial_x^q
&=\hat{x}^p\partial_x^q f(\hat{D}_x+p-q)\\
\hat{x}^p\partial_x^qf(\hat{D}_x)
&=f(\hat{D}_x-p+q)\hat{x}^p\partial_x^q\,
\end{align}
\end{subequations}
\begin{proof}
Via straightforward computations, applying the above operator formulae to monomials $x^n$.
\end{proof}
\end{Lemma}

As a first application, these auxiliary relations allow us to refine the formula for the monic SJ polynomials $\tilde{P}^{(-1,-1)}_n(x)$ as follows:
\begin{Proposition}
The monic Sobolev-Jacobi polynomials $\tilde{P}_n^{(-1,-1)}(x)$ are given in terms of the equivalent \emph{exponential formula}
\begin{equation}\label{eq:SJexp}
\tilde{P}_n^{(-1,-1)}(x)=e^{-\tfrac{1}{2}\tfrac{1}{\hat{D}_x+n-1}\partial_x^2}x^n\,
\end{equation}
Here, the operator in the denominator of~\eqref{eq:SJexp} is understood as rendered invertible via a suitable completion of its kernel without modifying the final result (compare Proposition~\ref{prop:SJ1}).
\begin{proof}
Starting from the original definition of $\tilde{P}^{(-1,-1)}_n(x)$ as given in Proposition~\ref{prop:SJ1}, the claim follows from the following application of Lemma~\ref{lem:ENO}:
\begin{align*}
\left[\tfrac{1}{(\hat{D}_x-n)(\hat{D}_x+n-1)}\partial_x^2\right]^mx^n &\overset{\eqref{eq:ENO1}}{=}\left[\tfrac{1}{(\hat{D}_x+n-1)}\partial_x^2\right]^m
\left(\prod_{j=1}^m\tfrac{1}{\hat{D}_x-n-2j}\right)x^n=\tfrac{1}{m!}\left(-\tfrac{1}{2}\right)^m\left[\tfrac{1}{(\hat{D}_x+n-1)}\partial_x^2\right]^mx^n\,
\end{align*}%
\end{proof}
\end{Proposition}

\subsection{Exponential Generating Function}
\label{sec:SJgf}

It will prove convenient to introduce the following bi-variate polynomials:
\begin{equation}\label{eq:PNXY}
P_n(x,y):=e^{\bB}(xy)^n\,
\end{equation}
where the differential operator $\bB$ is defined as
\begin{equation}\label{eq:defBB}
\bB:=\bfb_0 \,\partial^2_x\,,\quad \bfb_p:=-\tfrac{1}{2}\tfrac{1}{\hat{D}_x+\hat{D}_y+p-1}\quad (p\in \bZ_{\geq 0})\,
\end{equation}

As in previous formulae involving operators in denominators, we tacitly employ the same strategy of completing the kernels of these operators by any invertible function such as the identity function on their kernel, thereby rendering them invertible. This entails that in concrete applications, one must check that the resulting formulae are \emph{independent} of these choices of completions.

Returning to the definition of the polynomials $P_n(x,y)$ as given in~\eqref{eq:PNXY}, we evidently have that
\[
\tilde{P}^{(-1,-1)}_n(x)=\left(P_n(x,y)\right)\bigg\vert_{y\to1}\,
\]

To compute the EGF of the SJ polynomials $\tilde{P}^{(-1,-1)}_n(x)$, we~first compute the one of the polynomials $P_n(x,y)$---in practice, the fact that the differential operator $\bB$ does not explicitly depend on the degree $n$ yields a computational advantage, as the following result~demonstrates:
\begin{Proposition}\label{prop:SJexp}
The EGF $\cG(\lambda;x,y)$ of the polynomials $P_n(x,y)$, defined as
\begin{equation}
\cG(\lambda;x,y):=\sum_{n\geq 0}\frac{\lambda^n}{n!}P_n(x,y)=e^{\bB}e^{\lambda xy}
\end{equation}
has the explicit form
\begin{subequations}\label{eq:SJegf1}
\begin{align}
\cG(\lambda;x,y)
&=\sum_{n\geq 0}\tfrac{(\lambda x y)^n}{n!}\sum_{m\geq 0}\tfrac{1}{m!}\left(-\tfrac{\lambda^2 y^2}{4}\right)^m\tfrac{\Gamma(m+n-\tfrac{1}{2})}{\Gamma(2m+n-\tfrac{1}{2})}\\
&=\sum_{n\geq 0}\frac{(\lambda x y)^n}{n!}\; \pFq{1}{2}{n-\tfrac{1}{2}}{\tfrac{n}{2}-\tfrac{1}{4},\tfrac{n}{2}+\tfrac{1}{4}}{-\tfrac{\lambda^2y^2}{16}}\,
\end{align}
\end{subequations}
which is one of the key results of this paper.
\begin{proof}
See Appendix~\ref{app:SJgen}.
\end{proof}
\end{Proposition}

\subsection{Series Transform Techniques and Sobolev-Jacobi Polynomials}
\label{sec:SJSTT}

Upon closer inspection, the formula for the EGF of the SJ polynomials as presented in~\eqref{eq:SJegf1} contains in particular a fraction of gamma functions. Expressing these gamma functions via our integral operator $\tio$ reveals a hidden structure in $\cG(\lambda;x,y)$:
\begin{subequations}
\begin{align}
\cG(\lambda;x,y)
&=\sum_{n\geq 0}\tfrac{(\lambda x y)^n}{n!}\sum_{m\geq 0}\tfrac{1}{m!}\left(-\tfrac{\lambda^2 y^2}{4}\right)^m\tfrac{\Gamma(m+n-\tfrac{1}{2})}{\Gamma(2m+n-\tfrac{1}{2})}\\
&=\tio\left(
\sum_{n\geq 0}\tfrac{(\lambda x y)^n}{n!}\sum_{m\geq 0}\tfrac{1}{m!}\left(-\tfrac{\lambda^2 y^2}{4}\right)^m
u^{m+n-\tfrac{1}{2}}v^{2m+n-\tfrac{1}{2}}
\right)\\
&=\tio\left(
\frac{1}{\sqrt{uv}}\left(\sum_{n\geq 0}\frac{(\lambda u v x y)^n}{n!}\right)
\left(\sum_{m\geq 0}\frac{1}{m!}\left(-\frac{\lambda^2 u v^2 y^2}{4}\right)^n\right)
\right)\,
\end{align}
\end{subequations}
whence
\begin{equation}\label{eq:SJgfA}
\cG(\lambda;x,y)=\tio\left(
\frac{1}{\sqrt{uv}} e^{\lambda (uvxy)+\lambda^2 \left(-\tfrac{uv^2y^2}{4}\right)}
\right)\,
\end{equation}
To proceed, let us recall (see e.g., \cite{dattoli2000note}) the definition of the bi-variate Hermite polynomials $H_r(x,z)$,
\begin{equation}\label{eq:BH}
H_r(x,z)=r!\sum_{m=0}^{\lfloor\tfrac{r}{2}\rfloor}\tfrac{x^{r-2m}z^m}{m!(r-2m)!}\,
\end{equation}
as well as their well-known \emph{EGF}
\begin{equation}\label{eq:EGFhp}
\cH(\lambda;x,z):=\sum_{r\geq 0}\tfrac{\nu^r}{r!}H_r(x,z)=e^{\nu x+\nu^2 z}\,
\end{equation}
which may be derived from the important operational identity~\cite{dattoli1997evolution,dattoli1999operational}
\begin{equation}
H_n(x,z)=e^{z\partial^2_x}x^n\,
\end{equation}
Here, to avoid any potential notational confusion, we have denoted the second variable of the Hermite polynomials $H_n(x,z)$ by $z$ (rather than $y$), since in some of our formulae the notation $y$ is used in a different context (i.e., for defining the SJ~polynomials).

Another useful set of identities is given by
\begin{equation}
\begin{aligned}\label{eq:Heqs}
\partial_x^2 H_n(x,y)&=\partial_y H_n(x,y)=(n)_2 H_{n-2}(x,y)\,
\end{aligned}
\end{equation}
with these preparations, we find the following result with far-reaching consequences for the rest of this paper, and which hints at the potential of our novel series transformation methods:
\begin{Theorem}\label{thm:SJP}
The EGF $\cG(\lambda;x)$ is an integral-based series transform of the generating function $\cH(\lambda;x,z)$,
\begin{equation}\label{eq:SJegfUF}
\cG(\lambda;x)=\tio\left(
\frac{1}{\sqrt{uv}} e^{(\lambda uv)x+(\lambda uv)^2\left(-\tfrac{1}{4u}\right)}
\right)
=\tio\left(
\frac{1}{\sqrt{uv}} \cH\left(\lambda uv; x,-\tfrac{1}{4u}\right)
\right)\,
\end{equation}

This entails the following explicit formula for the monic SJ polynomials $P_n(x)\equiv \tilde{P}^{(-1,-1)}_n(x):=P_n(x,1)$:
\begin{equation}\label{eq:monSJ1}
\begin{aligned}
P_n(x)\equiv\tilde{P}^{(-1,-1)}_n(x)
&=\tio\left(
(uv)^{n-\tfrac{1}{2}}H_n(x,-\tfrac{1}{4u})
\right)=\tio\left(\frac{1}{\sqrt{uv}}e^{-\tfrac{1}{4u}\partial^2_x}(xuv)^n\right)\,
\end{aligned}
\end{equation}
\end{Theorem}

In practice, employing the definitions of Section~\ref{sec:ST}, Formula~\eqref{eq:monSJ1} allows for an efficient computation of the explicit forms of the SJ polynomials. The evaluated explicit formulae for the first few polynomials are presented in Table~\ref{tab:SJ1}.

\subsection{Aside: General Operational Definitions for Calculating Lacunary Generating Functions}
\label{sec:LGFgen}

Suppose we were given the EGF $G(\lambda;x)$ of some set of polynomials $p_n(x)$ (where $n\in \bZ_{\geq 0}$, and where we assume $degree(p_n(x))=n$). We will be particularly interested in the relationship between the following two forms of presenting $G(\lambda;x)$ as a formal power series:
\begin{subequations}
\begin{align}
G(\lambda;x)&=\sum_{n\geq 0}\frac{\lambda^n}{n!} p_n(x)\label{eq:GFformA}\\
&=\sum_{n\geq 0}x^n\sum_{m\geq 0} \frac{\lambda^{m+n}}{n!} g_{n,m}\,\label{eq:GFformB}
\end{align}
\end{subequations}

In certain situations, one might be interested in so-called \emph{lacunary generating functions}:
\begin{Definition}
Given an EGF $G(\lambda;x)=\sum_{n\geq 0}\frac{\lambda^n}{n!}p_n(x)$, the \emph{$K$-tuple $L$-shifted lacunary generating function} $G_{K,L}(\lambda;x)$ of the polynomials $p_n(x)$ is defined as (for $K\in \bZ_{\geq 1}$ and $L\in \bZ_{\geq 0}$)
\begin{equation}
G_{K,L}:=\sum_{n\geq 0}\frac{\lambda^n}{n!} p_{K\cdot n+L}(x)\,
\end{equation}\label{eq:defLGF}

Thus, in particular $G_{1,0}(\lambda;x)=G(\lambda;x)$.
\end{Definition}

The difficulty in explicit calculations of lacunary generating functions resides in finding the explicit coefficient form as indicated in~\eqref{eq:GFformB}. The series transform methods presented thus far allow to cast these computations into a concise operational form:
\begin{Lemma}
Let $G(\lambda;x)$ be an EGF, $K\in \bZ_{\geq 1}$ and $L\in \bZ_{\geq 0}$ integer parameters and
\begin{equation}
\eta_K:=e^{\frac{2\pi \mathbf{i}}{K}}
\end{equation}
a $K$-th root of unity ($(\eta_K)^K=1$). Then the lacunary generating function $G_{K,L}(\lambda;x)$ is given by
\begin{equation}\label{eq:lemLGF}
\begin{aligned}
G_{K,L}(\lambda;x)&=\LDO{K}\circ \left(\tfrac{\partial}{\partial\lambda}\right)^{L} G(\lambda;x)
\end{aligned}
\end{equation}
here, the \emph{lacunary dilatation operator} $\LDO{K}$ is defined as acting on a generic formal power series $F(\lambda;x)$ as
\begin{equation}\label{eq:defLD}
\LDO{K}F(\lambda;x):=\tio\left(\frac{uv}{k}\sum_{j=0}^{K-1}F\left(\eta_K^j u(v\lambda)^{\frac{1}{K}};x\right)\right)\,
\end{equation}

Moreover, the following relations hold:
\begin{equation}\label{eq:opsRel}
\left(\tfrac{\partial}{\partial\lambda}\right)^{L}\circ \LDO{K}=\LDO{K}\circ \left(\tfrac{\partial}{\partial\lambda}\right)^{K\cdot L}\,
\end{equation}
\begin{proof}
The first (well-known) part of the claim follows from
\begin{equation}
\begin{aligned}
\left(\tfrac{\partial}{\partial\lambda}\right)^{L} G(\lambda;x)=\sum_{n\geq L}\frac{\lambda^{n-L}}{(n-L)!}p_{n}(x)=\sum_{n\geq 0}\frac{\lambda^n}{n!}p_{n+L}(x)=G_{1,L}(\lambda;x)\,
\end{aligned}
\end{equation}

To prove the formula for the action of $\LDO{K}$ on a generic formal power series $F(\lambda;x)$, note first that for $k>1$
\begin{equation}
\begin{aligned}
\frac{1}{K}\sum_{j=0}^{K-1}(\eta_K^n)^j
& =\begin{cases}
\frac{1}{K}\sum_{j=0}^{K-1} 1\quad &\text{if } n\Mod K=0\\
\frac{1}{K}\frac{\eta_K^K-1}{\eta_K-1}&\text{else}
\end{cases}= \delta_{(n\Mod K),0}\,
\end{aligned}
\end{equation}
and that by definition for all $K\in \bZ_{\geq 1}$ and $n\in \bZ_{\geq 0}$,
\begin{equation}
\tio\left(uv(uv^{\frac{1}{K}})^n\right)=\frac{\Gamma(n+1)}{\Gamma(\frac{n}{K}+1)}\,
\end{equation}

Whence for an EGF $G(\lambda;x)=\sum_{n\geq 0}\frac{\lambda^n}{n!}p_n(x)$, the application of $\LDO{K}$ results in
\begin{equation}
\begin{aligned}
\LDO{K}G(\lambda;x)&=\tio\left(
\sum_{n\geq 0}\frac{\lambda^n}{n!}p_n(x)u^{n+1}v^{\frac{1}{K}+1}\frac{1}{K}\sum_{j=0}^{K-1}
\left(\eta_K^n\right)^j\right)\\
&=\sum_{n\geq 0}\frac{\lambda^n}{n!}p_n(x)\frac{\Gamma(n+1)}{\Gamma(\frac{n}{K}+1)}\delta_{(n\Mod K),0}\\
&=\sum_{r\geq 0}\frac{\lambda^r}{r!}p_{r\cdot K}(x)=\cG_{K,0}(\lambda;x)\,
\end{aligned}
\end{equation}
Finally, the relations presented~\eqref{eq:opsRel} follow directly from the definitions.
\end{proof}
\end{Lemma}

The readers will notice the similarity of the definition given in~\eqref{eq:defLD} with the technique of so-called \emph{multisection of series}~\cite{wilf2005generatingfunctionology}. However, the presence of the operator $\tio$ enriches this technique in a substantial way. For comparison, we refer the interested readers to our recent work~\cite{bdp2018hp} for an alternative, but~equivalent definition of the lacunary dilatation operator $\LDO{K}$.

\subsection{Lacunary Exponential Generating Functions for the SJ Polynomials for the Case $\alpha=\beta=-1$}
\label{sec:LGSJ}

While it may not be immediately evident, the true merit of the definition of the lacunary dilatation and shift operators $\LDO{K}$ and $\partial_{\lambda}$ is revealed upon applying them to computations of the EGF forms described in~\eqref{eq:GFformB} for the case of an EGF $G(\lambda;x)$. Some of the authors of the present paper~\cite{bdp2018hp} recently applied these ideas to the computation of explicit formulae for all lacunary generating functions $\cH_{K,L}(\lambda;x,z)$ (with $K\in \bZ_{\geq 1}$ and $L\in \bZ_{\geq 0}$) of the two-variable Hermite polynomials $H_n(x,y)$. We~quote the respective results in Appendix~\ref{app:HPlgf} for the interested readers' convenience.

For the present paper, we would like to focus on the interaction of series transforms with generating function formulae. More concretely, we will present an immediate consequence of this general theorem for the lacunary generating functions of the Hermite polynomials in terms of the corresponding generating functions of the SJ polynomials. As a tool that we will frequently invoke in the following considerations, let us recall the \emph{Gauss-Legendre multiplication formula}~(\cite{DLMF}, Equation~(5.5.6)) for Gamma functions (for $n\cdot z\not\in \bZ_{\leq 0}$),
\begin{equation}\label{eq:GMF}
\Gamma(nz)=n^{nz-\frac{1}{2}}(2\pi)^{\frac{(1-n)}{2}}\prod_{j=0}^{n-1}\Gamma\left(z+\frac{j}{n}\right)\,
\end{equation}

More precisely, we will make use of the following variant of this formula: for $n(s+x)$, \mbox{$nx\in \bC\setminus \bZ_{\leq 0}$}, $n\in \bZ_{\geq 2}$ and $s\in \bZ_{\geq 0}$, we have that
\begin{equation}\label{eq:GMFC}
\Gamma(n(s+x))=\left(n^{s\cdot n}\right)\Gamma(nx)\;\prod_{j=0}^{n-1}\left(x+\frac{j}{n}\right)_s\,
\end{equation}
with the Pochhammer symbol $(a)_b$ defined according to the convention
\[
(a)_b:=\frac{\Gamma(a+b)}{\Gamma(a)}\,
\]

To compute the lacunary generating functions $\cG_{K,L}(\lambda;x)$ of the SJ polynomials $\tilde{P}^{(-1,-1)}_n(x)$, we~merely have to apply Theorem~\ref{thm:SJP} to the results presented in Theorem~\ref{thm:lacHP}.~More~precisely, introducing~the notations $h^{(K,L)}_{r,m}(z)$ for the expansion coefficients of the generating functions $\cH_{K,L}(\lambda;x,z)$ and $g^{(K,L)}_{r,m}(z)$ of the generating functions $\cG_{K,L}(\lambda;x)$, respectively,
\begin{equation}\label{eq:defLacGFcoeffs}
\begin{aligned}
\cH_{K,L}(\lambda;x,z)&=\sum_{r\geq 0} x^r \sum_{m\geq 0} \frac{\lambda^{r+m}}{(r+m)!}\; h^{(K,L)}_{r,m}(z)\\
\cG_{K,L}(\lambda;x)&=\sum_{r\geq 0} x^r \sum_{m\geq 0} \frac{\lambda^{r+m}}{(r+m)!}\; g^{(K,L)}_{r,m}\,
\end{aligned}
\end{equation}
we find the following precise relationship between the generating functions $\cH_{K,L}(\lambda;x)$ and $\cG_{K,L}(\lambda;x,z)$:
\begin{Corollary}\label{cor:LGFhpToSJ}
For all $K\in \bZ_{\geq 1}$ and $L\in \bZ_{\geq 0}$, the lacunary generating functions $\cG_{K,L}(\lambda;x)$ of the SJ polynomials $P_n(x)\equiv\tilde{P}^{(-1,-1)}_n(x)$ and $\cH_{K,L}(\lambda;x,z)$ of the Hermite polynomials $H_n(x,z)$ are related according to
\begin{equation}\label{eq:LGFhpToSJ}
\cG_{K,L}(\lambda;x)=\tio\left(
(uv)^{L-\frac{1}{2}}\;\cH_{K,L}\left(\lambda(uv)^K; x,-\tfrac{1}{4u}\right)\right)\,
\end{equation}
On the level of expansion coefficients, this entails that
\begin{equation}\label{eq:EChpToSJ}
g^{(K,L)}_{r,m}=\tio\left(
(uv)^{K\cdot(r+m)+L-\frac{1}{2}}\; h^{(K,L)}_{r,m}\left(-\tfrac{1}{4u}\right)
\right)\,
\end{equation}
\end{Corollary}
\begin{proof}
The proof follows from inserting the expression given for $P_n(X)$ in~\eqref{eq:monSJ1} into the defining equation of the lacunary generating function $\cG_{K,L}(\lambda;x)$, and for analogously for the expansion coefficients as specified in~\eqref{eq:defLacGFcoeffs}.
\end{proof}

Finally, we will need the following auxiliary statement (throughout which we will consider $z$ as a \emph{formal} variable to avoid any issues of convergence, compare~(\cite{DLMF}, §16.2), according to which in particular all generalized hypergeometric series with $p>q+1$ are divergent):
\begin{Lemma}[``Pochhammer proliferation'']\label{lem:PPL}
For parameters $\alpha,\beta\in \bC\setminus\bZ_{\leq 0}$ and $r,s\in \bZ_{\geq 1}$, the following series transformation yields in effect a ``proliferation of Pochhammer symbols'':
\begin{equation}\label{eq:PPL}
\begin{aligned}
&\tio\left(
u^{\alpha}v^{\beta} \pFq{p}{q}{\seq{a_j}{1\leq j\leq p}}{\seq{b_{\ell}}{1\leq \ell\leq q}}{zu^rv^s}
\right)
=\frac{\Gamma(\alpha)}{\Gamma(\beta)}\;  \pFq{p+r}{q+s}{\left(\seq{a_j}{1\leq j\leq p},
\seq{\tfrac{\alpha+k}{r}}{0\leq k\leq r-1}\right)}{\left(\seq{b_{\ell}}{1\leq \ell\leq q},
\seq{\tfrac{\beta+t}{s}}{0\leq t\leq s-1}\right)}{\frac{z r^r}{s^s}}\,
\end{aligned}
\end{equation}
\end{Lemma}
\begin{proof}
The proof follows via a straightforward application of \eqref{eq:GMFC}:
{\allowdisplaybreaks
\begin{align*}
\tio\left(
u^{\alpha}v^{\beta} \pFq{p}{q}{\seq{a_j}{1\leq j\leq p}}{\seq{b_{\ell}}{1\leq \ell\leq q}}{zu^rv^s}
\right)&=\sum_{m\geq 0}\frac{z^m}{m!}\; \frac{\left(\prod_{j=1}^p (a_j)_m\right)}{\left(\prod_{\ell=1}^q (b_q)_m\right)}\;
\tio\left( u^{\alpha+m\cdot r}v^{\beta+m\cdot s}\right)\\
&=\sum_{m\geq 0}\frac{z^m}{m!}\; \frac{\left(\prod_{j=1}^p (a_j)_m\right)}{\left(\prod_{\ell=1}^q (b_q)_m\right)}\;
\frac{\Gamma(m\cdot r +\alpha)}{\Gamma(m\cdot s+\beta)}\\
&=\sum_{m\geq 0}\frac{z^m}{m!}\; \frac{\left(\prod_{j=1}^p (a_j)_m\right)}{\left(\prod_{\ell=1}^q (b_q)_m\right)}\;
\frac{\Gamma(\alpha)\;(\alpha)_{m\cdot r}}{\Gamma(\beta)\; (\beta)_{m\cdot s}}\\
&\overset{\eqref{eq:GMFC}}{=}
\sum_{m\geq 0}\frac{z^m}{m!}\; \frac{\left(\prod_{j=1}^p (a_j)_m\right)}{\left(\prod_{\ell=1}^q (b_q)_m\right)}\;
\frac{\Gamma(\alpha)}{\Gamma(\beta)}\;
\frac{(r^r)^m\; \prod_{k=0}^{r-1}\left(\tfrac{\alpha+k}{r}\right)_m}{(s^s)^m\; \prod_{t=0}^{s-1}\left(\tfrac{\beta+t}{s}\right)_{m}}\\
&=\frac{\Gamma(\alpha)}{\Gamma(\beta)}\;  \pFq{p+r}{q+s}{\left(\seq{a_j}{1\leq j\leq p},
\seq{\tfrac{\alpha+k}{r}}{0\leq k\leq r-1}\right)}{\left(\seq{b_{\ell}}{1\leq \ell\leq q},
\seq{\tfrac{\beta+t}{s}}{0\leq t\leq s-1}\right)}{\frac{z r^r}{s^s}}\,
\end{align*}}
\end{proof}

Making use of the results for the lacunary generating functions of the Hermite polynomials $H_n(x,y)$ and of all of the corollaries and lemmata presented thus far, we are finally in a position to derive the following closed-form results for all lacunary generating functions of the SJ polynomials $P_n(x)\equiv\tilde{P}^{(-1,-1)}_n(x)$. To the best of our knowledge, all these formulae appear to be new.

\begin{Theorem}[All-order lacunary generating functions for the SJ polynomials]\label{thm:lacAll}
For integer parameters $K\in \bZ_{\geq 1}$, denote by
\begin{equation}\label{eqDefGK}
\cG_K(\mu;\lambda;x):=\sum_{L\geq 0}\frac{\mu^L}{L!}\;\cG_{K,L}(\lambda;x)
\end{equation}
the EGF of lacunary shifts of the lacunary generating functions $\cG_{K,0}(\lambda;x)$. Then~$\cG_K(\mu;\lambda;x)$ is given by
\begin{equation}\label{eq:SJlgfS}
\cG_K(\mu;\lambda;x)=\tio\left(
\tfrac{1}{\sqrt{uv}}\; e^{\mu uvx-\frac{\mu^2 u v^2}{4}}\; \cH_{K,0}\left(\lambda(uv)^K;x-\tfrac{\mu v}{2},-\tfrac{1}{4u}\right)\right)\,
\end{equation}
which permits (making use of ``Pochhammer proliferation'', Lemma~\ref{lem:PPL} and the explicit formulae for $\cH_{k,0}(\lambda;x,y)$ presented in Theorem~\ref{thm:lacHP}) to calculate arbitrary $L$-fold lacunary shifted generating functions $\cG_{K,L}(\lambda;x)$. In~particular, the explicit formulae for $L=0$ read in the case $K=2T$ (for $T\in \bZ_{\geq 1}$)
\begin{equation}\label{eq:LGFsjEven}
\begin{aligned}
\cG_{K=2T,0}(\lambda;x)&=
\sum_{\beta=0}^{T-1}\sum_{s\geq 0}
\tfrac{\lambda^s}{s!}
x^{K\cdot s-2\beta}\left(-\tfrac{1}{4}\right)^{\beta}\tilde{h}_{K\cdot s,\beta}\tfrac{\Gamma(K\cdot s-\beta-\tfrac{1}{2})}{\Gamma(K\cdot s-\tfrac{1}{2})}\;\cdot\\
&\qquad
\pFq{(3T-1)}{(3T-1)}{%
\left(\seq{s+\tfrac{j+1}{K}}{0\leq j\leq K-2},
\seq{2s+\tfrac{2m-\beta-1}{K}}{0\leq m\leq T-1}\right)}{%
\left(
\seq{\tfrac{\beta+\ell+1}{T}}{\substack{0\leq \ell\leq T-1\\ \ell\neq T-1-\beta}},
\seq{s+\tfrac{2t-1}{2K}}{0\leq t\leq K-1}
\right)}{\lambda\left(-\tfrac{1}{4}\right)^T}
\end{aligned}
\end{equation}
and in the case $K=2T+1$ (for $T\in \bZ_{\geq 0}$)
\begin{myequation}\label{eq:LGFsjOdd}
\begin{aligned}
\cG_{K=2T+1,0}(\lambda;x)&=
\sum_{\beta=1}^{K-1}\sum_{s\geq0}
\tfrac{\lambda^s}{s!}x^{K\cdot s-2\beta}
\left(-\tfrac{1}{4}\right)^{\beta}\tilde{h}_{K\cdot s,\beta}
\tfrac{\Gamma(K\cdot s-\beta-\tfrac{1}{2})}{\Gamma(K\cdot s-\tfrac{1}{2})}\; \cdot\\
&\qquad
\pFq{(3K-2)}{(3K-1)}{\left(
\left(\tfrac{s}{2}+\tfrac{j+1}{2K}\right)_{\substack{0\leq j \leq 2K-2\\ j\neq K-1}},
\left(s+\tfrac{2m-2\beta-1}{2K}\right)_{0\leq m\leq K-1}
\right)}{\left(
\left(\tfrac{\beta+\ell+1}{K}\right)_{\substack{0\leq \ell \leq K-1\\ \ell\neq K-1-\beta}},
\left(\tfrac{s}{2}+\tfrac{2t-1}{4K}\right)_{0\leq t\leq 2K-1}
\right)}{-\frac{\lambda^2}{4^{K+1}}}\,
\end{aligned}
\end{myequation}
here, $\tilde{h}_{n,k}$ denote the so-called \emph{matching coefficients} (of directed Hermite-configurations),
\begin{equation}\label{eq:HPmatch}
\tilde{h}_{n,k}:=\begin{cases}
\frac{n!}{(n-2k)!k!} &\qquad \text{if $0\leq 2k\leq n$}\\
0 &\qquad \text{otherwise}
\end{cases}
\end{equation}
\end{Theorem}
\begin{proof}

While~\eqref{eq:LGFsjEven} and~\eqref{eq:LGFsjOdd} follow directly from Theorem~\ref{thm:lacHP}, Corollary~\ref{cor:LGFhpToSJ} and Lemma~\ref{lem:PPL}, it remains to prove~\eqref{eq:SJlgfS}, whence the formula for the generating functions $\cG_{K}(\mu;\lambda;x)$ of the $L$-fold lacunary shifts of the $K$-tuple lacunary generating functions $\cG_{K,L}(\lambda;x)$. Recall from Appendix~\ref{app:HPlgf}, Equation~\eqref{eq:propHLGF} the corresponding formula for the Hermite polynomials (where for later convenience we rename the formal variable that was denoted $\mu$ in~\eqref{eq:propHLGF} to $\sigma$),
\begin{equation}
\cH_K(\sigma;\lambda;x,z):= \sum_{L\geq 0}\frac{\sigma^L}{L!} \cH_{K,L}(\lambda;x,z)=e^{\sigma x+\sigma^2 z}\cH_{K,0}(\lambda;x+2\sigma z,z)\,
\end{equation}

Since accordingly
\begin{equation}\label{eq:proofAuxA}
\begin{aligned}
\cH_{K,L}(\lambda;x,z)&=\left[\frac{\partial^L}{\partial \sigma^L}\cH_K(\sigma;\lambda;x,z)\right]\bigg\vert_{\sigma\to 0}\\
&=\left[\frac{\partial^L}{\partial \sigma^L}e^{\sigma x+\sigma^2 z}\cH_{K,0}(\lambda;x+2\sigma z,z)\right]\bigg\vert_{\sigma\to 0}\,
\end{aligned}
\end{equation}

Combining this result with the result of~\eqref{eq:LGFhpToSJ} on the relationship of $\cG_{K,L}(\lambda;x)$ and $\cH_{K,L}(\lambda;x,z)$ into the defining Equation~\eqref{eqDefGK} for  $\cG_K(\mu;\lambda;x)$, we obtain
\begin{equation}
\begin{aligned}
\cG_K(\mu;\lambda;x)&\overset{\eqref{eqDefGK}}{=}\sum_{L\geq 0}\frac{\mu^L}{L!}\; \cG_{K,L}(\lambda;x)
\overset{\eqref{eq:LGFhpToSJ}}{=}\sum_{L\geq 0}\frac{\mu^L}{L!}\;\tio\left(
(uv)^{L-\frac{1}{2}}\;\cH_{K,L}\left(\lambda(uv)^K; x,-\tfrac{1}{4u}\right)\right)\\
&\overset{\eqref{eq:proofAuxA}}{=}
\tio\left(\frac{1}{\sqrt{uv}}\; \left[\sum_{L\geq 0}\frac{(\mu uv)^L}{L!}\frac{\partial^L}{\partial \sigma^L}
e^{\sigma x-\frac{\sigma^2}{4u}}\cH_{K,0}(\lambda (uv)^K;x-\tfrac{\sigma}{2u},-\tfrac{1}{4u})\right]\bigg\vert_{\sigma\to 0}\right)\\
&=
\tio\left(\frac{1}{\sqrt{uv}}\; \left[e^{(\mu uv)\frac{\partial}{\partial \sigma}}\left(
e^{\sigma x-\frac{\sigma^2}{4u}}\cH_{K,0}(\lambda (uv)^K;x-\tfrac{\sigma}{2u},-\tfrac{1}{4u})\right)\right]\bigg\vert_{\sigma\to 0}\right)\\
&\overset{(*)}{=}
\tio\left(\frac{1}{\sqrt{uv}}\; \left[
e^{(\sigma+\mu uv) x-\frac{(\sigma+\mu uv)^2}{4u}}
\cH_{K,0}(\lambda (uv)^K;x-\tfrac{(\sigma+\mu uv)}{2u},-\tfrac{1}{4u})\right]\bigg\vert_{\sigma\to 0}\right)\\
&=\tio\left(
\tfrac{1}{\sqrt{uv}}\; e^{\mu uvx-\frac{\mu^2 u v^2}{4}}\; \cH_{K,0}\left(\lambda(uv)^K;x-\tfrac{\mu v}{2},-\tfrac{1}{4u}\right)\right)\,
\end{aligned}
\end{equation}
here, in the step marked $(*)$, we have made use of Taylor's formula, $e^{\alpha\frac{\partial}{\partial \sigma}}f(\sigma)=f(\sigma+\alpha)$.
\end{proof}

For illustration, we present several examples for lacunary generating functions $\cG_{K,L}(\lambda;x,y)$ in~Table~\ref{tab:SJlgf}.

\subsection{Connection Coefficients for SJ Polynomials for the Case $\alpha=\beta=-1$}
\label{sec:CCSJ}

Consider the so-called \emph{connection coefficients} $A_{M,n}$ of the SJ polynomials $\tilde{P}^{(-1,-1)}_n(x)$ (see~\cite{bdp2017} for an explicit solution in terms of integral computations), which are defined for non-negative integers $M$ and $n$ as
\begin{equation}\label{eq:AMN}
x^M=\sum_{n=0}^M A_{M,n} \tilde{P}^{(-1,-1)}_n(x)\,
\end{equation}

In this section, we will highlight some aspects of computations related to these coefficients in terms of operational techniques. To this end, we first derive the connection coefficients for the two-variable Hermite polynomials $H_n(x,y)$~\cite{dattoli2000generalized}:
\begin{equation}
\begin{aligned}
H_M(x,y)&=e^{y\partial^2_x}x^M\\
\Rightarrow\quad
x^M&=e^{-y\partial^2_x}H_M(x,y)=\sum_{n=0}^{\lfloor \tfrac{M}{2}\rfloor} \frac{(-y)^n}{n!} \partial_x^{2n}H_M(x,y)\\
&\overset{\eqref{eq:Heqs}}{=}
\sum_{n=0}^{\lfloor \tfrac{M}{2}\rfloor} \frac{(-y)^n M!}{n! (M-2n)!} H_{M-2n}(x,y)\\
&\equiv\sum_{n\geq 0}a_{M,n}(y)H_n(x,y)\,
\end{aligned}
\end{equation}

In consequence, the generating function for the connection coefficients $a_{M,n}(y)$ of the polynomials $H_n(x,y)$ reads
\begin{equation}
\begin{aligned}
A_H(\lambda,\mu;y)&:=\sum_{M,n\geq 0}\frac{\lambda^M}{M!}\mu^n a_{M,n}(y)\\
&=\sum_{M\geq 0}\frac{\lambda^M}{M!}
\sum_{n=0}^{\lfloor \tfrac{M}{2}\rfloor} \frac{M!\mu^{M-2n}(-y)^n}{n! (M-2n)!}\\
&\overset{\eqref{eq:BH}}{=}
\sum_{M\geq 0}\frac{\lambda^M}{M!} H_M(\mu,-y)\\
&\overset{\eqref{eq:EGFhp}}{=}
e^{\lambda \mu-\lambda^2 y}\,
\end{aligned}
\end{equation}

For later convenience, let us also consider a consistency check for this result involving the EGF $B_H(x,y)$ for the two-variable Hermite polynomials,
\begin{equation}\label{eq:HGF}
\begin{aligned}
B_H(\lambda,\mu;y)&:=\sum_{n\geq 0}\frac{\lambda^n}{n!}H_n(\mu,y)\equiv
\sum_{n,M\geq 0}\frac{\lambda^n}{n!}\mu^M b_{n,M}(y)=e^{\lambda\mu +\lambda^2 y}\,
\end{aligned}
\end{equation}

Making use of the well-known integral formula (for $w$ a complex variable and $d^2w\equiv dwd\bar{w}$)
\begin{equation}
\tfrac{1}{\pi}\int d^2w\; e^{-|w|^2} w^r\bar{w}^s=r!\delta_{r,s}\,
\end{equation}
the generating functions $A_H(\lambda,\mu;y)$ and $B_H(\lambda,\mu;y)$ verify the following property:
\begin{equation}
\begin{aligned}
\tfrac{1}{\pi}\int d^w\;e^{-|W|^2} A_H(\alpha,w;y)B_H(\bar{w},\beta;y)=e^{\alpha \beta}\,
\end{aligned}
\end{equation}

This is of course nothing else but a direct consequence of the  defining property of the coefficients $a_{M,n}(y)$ and $b_{n,M}(y)$, namely
\begin{equation}
\begin{aligned}
x^M&=\sum_{n\geq 0}a_{M,n}(y) H_n(x,y)
=\sum_{n,L\geq 0} a_{M,n}(y)b_{n,L}(y)x^L\\
\Rightarrow\quad &\sum_{n\geq 0} a_{M,n}(y)b_{n,L}(y)=\delta_{M,L}\,
\end{aligned}
\end{equation}
with these preparations, let us quote from~\cite{bdp2017} the explicit formula for the connection coefficients $A_{M,n}$ of the SJ polynomials $P_n(x)\equiv \tilde{P}^{(-1,-1)}_n(x)$ (note however the change of notations compared to~\cite{bdp2017}, i.e., $A_{M,n}\equiv B^{M,n}_{-1}$):
\begin{equation}
A_{M,n}=\begin{cases}
\binom{2n}{n}\frac{M! \left(\tfrac{M+n}{2}\right)!}{(M+n)!\left(\tfrac{M-n}{2}\right)!}\quad &\text{if } M+n\in 2\mathbb{Z}\\
&\quad \text{and } n\leq M\\
0 &\text{else}
\end{cases}
\end{equation}

By analogy with the two-variable Hermite polynomial case, let us define the generating function $A(\lambda,\mu)$ of these coefficients. Making use of the identity
\[
\Gamma(m+\tfrac{1}{2})=\frac{\sqrt{\pi} (2m)!}{4^m m!}\,
\]
we first rewrite the coefficients $A_{M,M-2n}$ as
\begin{equation}
\begin{aligned}
A_{M,M-2n}&=\binom{2M-4n}{M-2n}\frac{M!(M-n)!}{(2(M-p))!p!}\\
&=\frac{M!}{n!(M-2n)!}\frac{\Gamma(M-2n+\tfrac{1}{2})}{4^n\Gamma(M-n+\tfrac{1}{2})}\\
&=\frac{M!}{n!(M-2n)!}\;\tio\left(
(uv)^{M-2n+\tfrac{1}{2}}\left(\tfrac{v}{4}\right)^n
\right)\,
\end{aligned}
\end{equation}

We thus obtain a result reminiscent of the two-variable Hermite polynomial case for the generating function of the connection coefficients $A_{M,n}$:
\begin{equation}
\begin{aligned}
A(\lambda,\mu)&:=\sum_{M,n\geq 0}\frac{\lambda^M}{M!}\mu^n A_{M,n}\\
&=
\tio\left(
\sqrt{uv}\; \sum_{M\geq 0}\frac{\lambda^M}{M!}H(\mu uv,\tfrac{v}{4})
\right)\\
&=\tio\left(
\sqrt{uv}\; e^{\lambda \mu uv+\lambda^2 \tfrac{v}{4}}
\right)
=\tio\left(
\sum_{M\geq 0}\frac{(\lambda\mu)^M}{M!} \; (uv)^{M+\frac{1}{2}}\; e^{\lambda^2 \tfrac{v}{4}}
\right)\\
&\overset{\eqref{eq:pFqSIT}}{=}\sum_{M\geq 0}\frac{(\lambda\mu)^M}{M!} {}_0F_1(M+\tfrac{1}{2};\tfrac{\lambda^2}{4})\,
\end{aligned}
\end{equation}

While we were not able to derive this result purely via operational methods, but only by making use of the explicit formula provided in~\cite{bdp2017} for the connection coefficients $A_{M,n}$, the operational methods will nonetheless allow us to provide an independent verification of this explicit formula. To~this end, introduce the notation (compare~\eqref{eq:SJegfUF})
\[
B(\lambda,\mu):=\cG(\lambda;\mu)
=\tio\left(
\frac{1}{\sqrt{uv}}e^{\lambda \mu uv+\lambda^2\left(-\tfrac{uv^2}{4}\right)}\right)\,
\]
\begin{restatable}{prop}{ABprop}\label{prop:AB}

The generating functions $A(\lambda,\mu)$ and $B(\lambda,\mu)$ verify
\begin{equation}
\frac{1}{\pi}\int d^2w\; e^{-|w|^2} A(\alpha,w)B(\bar{w},\beta)=e^{\alpha\beta}\,
\end{equation}
\end{restatable}
\begin{proof}
See Appendix~\ref{app:AB}.
\end{proof}

%%%%%%%%%%%%%%%%%%%%%%%%%%%%%%%%%%%%%%%%%%
\section{Conclusions}

In this paper, we have applied operational methods
to the study of SJ polynomials. We have implemented, starting from first principles, a variant of multi-variate umbral calculus in the form of a certain integral transform technique based on the notions of generalized monomials and on the elementary integrals defining the special functions $\Gamma(z)$ and $1/\Gamma(z)$ (cf. Section~\ref{sec:ST}). Our~approach provides a bridge between the plethora of results and specialized techniques known from the umbral calculus literature and more conventional mathematical techniques of elementary algebra and calculus, which we hope will allow a wider audience to profit from this fruitful approach.~To apply our technique to the study of SJ polynomials, we adapted a remarkable method invented by Gurappa and Panigrahi~\cite{gurappa1996new,gurappa1999equivalence,gurappa2001novel,gurappa2002linear,Shreecharan_2004,gurappa2004polynomial} (cf. Section~\ref{sec:SJP}) as well as a certain generalized normal-ordering technique (for expressions involving Euler operators and fractions thereof, cf. Section~\ref{sec:ENO}). We~have achieved not only a deeper understanding of the Sobolev-Jacobi polynomials for parameters \mbox{$\alpha=\beta=-1$} as \emph{umbral images} of the two-variable Hermite polynomials $H_n(x,y)$ (see Theorem~\ref{thm:SJP}), but~were in consequence in particular able to derive explicit formulae for \emph{all} $K$-tuple $L$-shifted lacunary exponential generating functions (see Theorem~\ref{thm:lacAll}) for these SJ polynomials (based on a previous work~\cite{bdp2018hp} on lacunary EGFs for the bi-variate Hermite polynomials). A crucial step in this derivation consisted of a technical Lemma called ``Pochhammer Proliferation Lemma'' (See Lemma~\ref{lem:PPL}), which~highlights the strengths of our integral transform techniques and fully explains the explicit structure of the lacunary EGFs as formal power series over certain generalized hypergeometric functions. To~the best of our knowledge, these~results have not been known before in the literature. Additional~results include an explicit formula for the connection coefficients of the aforementioned family of SJ polynomials (cf. Section~\ref{sec:CCSJ}), a derivation of the EGFs for the remaining family of SJ polynomials (cf. Section~\ref{sec:SJalt}), and finally several auxiliary results and examples (cf. Section~\ref{sec:ucEx}) that illustrate our novel umbral calculus-type integral transform techniques, and which highlight a certain resemblance with earlier ideas as expressed in~\cite{dattoli2017operational,licciardi2018umbral} (referred to therein as the ``principle'' of permanence of formal properties). The results we have obtained appear to be promising and seem to offer new and interesting developments within the framework of the theory of special functions and of the relevant applications. A forthcoming investigation, along the lines suggested here, will be devoted to further progress towards a unified theory of Legendre-like polynomials.

\vspace{6pt}

%%%%%%%%%%%%%%%%%%%%%%%%%%%%%%%%%%%%%%%%%%
\authorcontributions{Conceptualization, N.B.; Methodology, N.B., G.D., G.H.E.D., K.A.P.; Formal analysis, N.B.; Writing: Original Draft Preparation, N.B.; Writing: Review \& Editing, N.B., G.D., G.H.E.D., S.L., K.A.P.}

%%%%%%%%%%%%%%%%%%%%%%%%%%%%%%%%%%%%%%%%%%
\funding{The work of N.B. was supported by a {H2020 Marie Sk\l{}odowska-Curie Actions Individual Fellowship} (grant~\#~753750---RaSiR).}

%%%%%%%%%%%%%%%%%%%%%%%%%%%%%%%%%%%%%%%%%%
\acknowledgments{N.B. would like to thank the LPTMC (Paris 06) and ENEA Frascati for warm hospitality.}

%%%%%%%%%%%%%%%%%%%%%%%%%%%%%%%%%%%%%%%%%%
\conflictsofinterest{The authors declare no conflicts of interest. The founding sponsors had no role in the design of the study; in the collection, analyses, or interpretation of data; in the writing of the manuscript, and in the decision to publish the results.}

\appendixtitles{yes}
\appendix

\section{Details of the Derivation of the Sobolev-Jacobi Polynomials\label{app:SJdef}}

Following Gurappa and Panigrahi~\cite{gurappa1996new,gurappa1999equivalence,gurappa2001novel,gurappa2002linear,Shreecharan_2004,gurappa2004polynomial}, writing the defining eigenequation~\eqref{eq:JacEigen} in explicit form, one may collect the terms in the following form (note however the systematic typographical error in loc.\ cit.\ in the definition of the diagonal part, compare e.g., \cite{gurappa2004polynomial}):
\begin{equation}
\begin{aligned}
0&=(\hat{D}^{(\alpha,\beta)}_{Jac}-(-n(n+\alpha+\beta+1)))\tilde{P}^{(\alpha,\beta)}_n(x)\\
&=\left((1-\hat{x}^2)\partial_x^2
+(\beta-\alpha-(\alpha+\beta+2)x)\partial_x +n(n+\alpha+\beta+1)\right)\tilde{P}^{(\alpha,\beta)}_n(x)\\
&=\left(\left[
-\hat{x}^2\partial_x^2-(\alpha+\beta+2)\hat{x}\partial_x
+n(n+\alpha+\beta+1)
\right]
+\left[
\partial_x^2+(\beta-\alpha)\partial_x
\right]\right)\tilde{P}^{(\alpha,\beta)}_n(x)\\
&\overset{(*)}{=}\left(\left[
-\hat{D}_x^2+\hat{D}_x-(\alpha+\beta+2)\hat{D}_x
+n(n+\alpha+\beta+1)
\right]
+\left[
\partial_x^2+(\beta-\alpha)\partial_x
\right]\right)\tilde{P}^{(\alpha,\beta)}_n(x)\\
&=\left(\left[
-(\hat{D}_x+n+\alpha+\beta+1)\hat{D}_x+n\hat{D}_x
+n(n+\alpha+\beta+1)
\right]
+\left[
\partial_x^2+(\beta-\alpha)\partial_x
\right]\right)\tilde{P}^{(\alpha,\beta)}_n(x)\\
&=\left(\left[
-(\hat{D}_x+n+\alpha+\beta+1)(\hat{D}_x-n)
\right]
+\left[
\partial_x^2+(\beta-\alpha)\partial_x
\right]\right)\tilde{P}^{(\alpha,\beta)}_n(x)\\
&=(F^{(\alpha,\beta)}_n(\hat{D}_x)+N(\hat{x},\partial_x))\tilde{P}^{(\alpha,\beta)}_n(x)\,
\end{aligned}
\end{equation}

In the step marked $(*)$, we made use of the definition $\hat{D}_x=\hat{x}\partial_x$ as well as of the auxiliary relation $\hat{x}^2\partial_x^2=\hat{D}_x^2-\hat{D}_x$ (see~\eqref{eq:EulerSq}).

\begin{table}[H]
\centering
\caption{Sobolev-Jacobi polynomials $\tilde{P}^{(-1,-1)}_n(x)$ and bi-variate Hermite polynomials $H_n(x,y)$.  \label{tab:SJ1}}

%% \renewcommand{\arraystretch}{1.7}
%%  \begin{tabular*}{\textwidth}{@{\extracolsep{\fill}}LLL}
\begin{tabular}{ccc}
\toprule
\textbf{n} & \boldmath{$\tilde{P}^{(-1,-1)}_n(x)$} & \boldmath{$H_n(x,y)$}\\
\midrule
0 & 1 & 1 \\
1 & \emph{x} & \emph{x} \\
2 & $x^2-1$ & $x^2+2 y$ \\
3 & $x^3-x$ & $x^3+6 x y$ \\
4 & $x^4-\frac{6 x^2}{5}+\frac{1}{5}$ & $x^4+12
x^2 y+12 y^2$ \\
5 & $x^5-\frac{10 x^3}{7}+\frac{3 x}{7}$ &
$x^5+20 x^3 y+60 x y^2$ \\
6 & $x^6-\frac{5 x^4}{3}+\frac{5
x^2}{7}-\frac{1}{21}$ & $x^6+30 x^4 y+180 x^2
y^2+120 y^3$ \\
7 & $x^7-\frac{21 x^5}{11}+\frac{35
x^3}{33}-\frac{5 x}{33}$ & $x^7+42 x^5 y+420
x^3 y^2+840 x y^3$ \\
8 & $x^8-\frac{28 x^6}{13}+\frac{210
x^4}{143}-\frac{140 x^2}{429}+\frac{5}{429}$
& $x^8+56 x^6 y+840 x^4 y^2+3360 x^2 y^3+1680
y^4$ \\
9 & $x^9-\frac{12 x^7}{5}+\frac{126
x^5}{65}-\frac{84 x^3}{143}+\frac{7 x}{143}$
& $x^9+72 x^7 y+1512 x^5 y^2+10080 x^3
y^3+15120 x y^4$ \\
10 & $x^{10}-\frac{45 x^8}{17}+\frac{42
x^6}{17}-\frac{210 x^4}{221}+\frac{315
x^2}{2431}-\frac{7}{2431}$ & $x^{10}+90 x^8
y+2520 x^6 y^2+25200 x^4 y^3+75600 x^2
y^4+30240 y^5$ \\
\bottomrule
\end{tabular}
\end{table}
\unskip

%\begin{landscape}
% start a new page
\newpage
% change it to landscape
\paperwidth=\pdfpageheight
\paperheight=\pdfpagewidth
\pdfpageheight=\paperheight
\pdfpagewidth=\paperwidth
\newgeometry{layoutwidth=297mm,layoutheight=210 mm, left=2.7cm,right=2.7cm,top=1.8cm,bottom=1.5cm, includehead,includefoot}
\fancyheadoffset[LO,RE]{0cm}
\fancyheadoffset[RO,LE]{0cm}
%%%%%%%%%%%%%%%%%%%%%%%%%%%%%%%%%%%%%%%%%%%%%%%

\begin{longtable}{RL}
%\begin{table}[H]
%\begin{tabular}{cc}
\caption{\ Lacunary generating functions $\cG_{K,L}(\lambda;x)$ for the SJ polynomials $P_n(x)$ and $K=1,\dotsc,5$ and some selected values of $L\in \bZ_{\geq 0}$.\label{tab:SJlgf}}\tiny\\
%  \centering
\toprule\\
\cG_{K,0}(\lambda;x)&=\sum_{n\geq 0}\tfrac{\lambda^n}{n!} P_{n\cdot K}(x)\\[1.2em]
\midrule\\
\endfirsthead
\multicolumn{2}{c}{{\textbf{Table A2.} \textit{Cont.}}} \\
\toprule\\
\cG_{K,0}(\lambda;x)&=\sum_{n\geq 0}\tfrac{\lambda^n}{n!} P_{n\cdot K}(x)\\[1.2em]
\midrule\\
\endhead
\bottomrule
%\multicolumn{2}{l}{\textit{Continued on next page}}
\\
\endfoot
\bottomrule\\
\endlastfoot
\cG_{1,0}(\lambda;x)
&=\sum _{s\geq 0}
\tfrac{\lambda^s}{s!}  x^s\;
\pFq{1}{2}{s-\tfrac{1}{2}}{%
\tfrac{s}{2}-\tfrac{1}{4},
\tfrac{s}{2}+\tfrac{1}{4}}{%
-\tfrac{\lambda ^2}{16}}\\[1.5em]
\cG_{2,0}(\lambda;x)&=\sum_{s=0}^{\infty} \tfrac{\lambda^s}{s!} \; x^{2s} \; \pFq{2}{2}{s+\frac{1}{2},2 s-\frac{1}{2}}{s-\frac{1}{4},s+\frac{1}{4}}{-\frac{\lambda }{4}}\\[1.5em]
\cG_{2,1}(\lambda;x)&=
\sum_{s\geq 0} \tfrac{\lambda^s}{s!}\;x^{2s+1}\;
\pFq{2}{2}{%
s+\tfrac{3}{2},
2s+\tfrac{1}{2}}{%
s+\tfrac{1}{4},
s+\tfrac{3}{4}}{%
-\frac{\lambda}{4}}\\[1.5em]
\\
\cG_{3,0}(\lambda;x)&=\sum_{s=0}^{\infty} \tfrac{\lambda^s}{s!} \; x^{3s} \; \pFq{7}{8}{\frac{s}{2}+\frac{1}{6},\frac{s}{2}+\frac{1}{3},\frac{s}{2}+\frac{2}{3},\frac{s}{2}+\frac{5}{6},s-\frac{1}{6},s+\frac{1}{6},s+\frac{1}{2}}{\frac{1}{3},\frac{2}{3},\frac{s}{2}-\frac{1}{12},\frac{s}{2}+\frac{1}{12},\frac{s}{2}+\frac{1}{4},\frac{s}{2}+\frac{5}{12},\frac{s}{2}+\frac{7}{12},\frac{s}{2}+\frac{3}{4}}{-\frac{\lambda ^2}{256}}\\[1.5em]
&\quad-\sum_{s=0}^{\infty} \tfrac{\lambda^{s+1}}{(s+1)!} \;x^{3 (s+1)-2}\;\left(\frac{\Gamma \left(3 (s+1)-\frac{3}{2}\right)}{4 \Gamma \left(3 (s+1)-\frac{1}{2}\right)}\right)\; \left(\tfrac{(3(s+1))!}{(3(s+1) -2)!}\right)\;\cdot\\[1.5em]
&\qquad\pFq{7}{8}{\frac{s+1}{2}+\frac{1}{6},\frac{s+1}{2}+\frac{1}{3},\frac{s+1}{2}+\frac{2}{3},\frac{s+1}{2}+\frac{5}{6},(s+1)-\frac{1}{2},(s+1)-\frac{1}{6},(s+1)+\frac{1}{6}}{\frac{2}{3},\frac{4}{3},\frac{s+1}{2}-\frac{1}{12},\frac{s+1}{2}+\frac{1}{12},\frac{s+1}{2}+\frac{1}{4},\frac{s+1}{2}+\frac{5}{12},\frac{s+1}{2}+\frac{7}{12},\frac{s+1}{2}+\frac{3}{4}}{-\frac{\lambda ^2}{256}}\\[1.5em]
&\quad+\sum_{s=0}^{\infty} \tfrac{\lambda^{s+2}}{(s+2)!} \;x^{3 (s+2)-4}\;\left(\frac{\Gamma \left(3 (s+2)-\frac{5}{2}\right)}{16 \Gamma \left(3 (s+2)-\frac{1}{2}\right)}\right)\; \left(\tfrac{(3(s+2))!}{2!(3(s+2) -4)!}\right)\;\cdot\\[1.5em]
&\qquad\pFq{7}{8}{\frac{s+2}{2}+\frac{1}{6},\frac{s+2}{2}+\frac{1}{3},\frac{s+2}{2}+\frac{2}{3},\frac{s+2}{2}+\frac{5}{6},(s+2)-\frac{5}{6},(s+2)-\frac{1}{2},(s+2)-\frac{1}{6}}{\frac{4}{3},\frac{5}{3},\frac{s+2}{2}-\frac{1}{12},\frac{s+2}{2}+\frac{1}{12},\frac{s+2}{2}+\frac{1}{4},\frac{s+2}{2}+\frac{5}{12},\frac{s+2}{2}+\frac{7}{12},\frac{s+2}{2}+\frac{3}{4}}{-\frac{\lambda ^2}{256}}\\[1.5em]
\\
\\
\cG_{4,0}(\lambda;x)&=
\sum_{s=0}^{\infty} \tfrac{\lambda^s}{s!} \; x^{4s} \; \pFq{5}{5}{s+\frac{1}{4},s+\frac{1}{2},s+\frac{3}{4},2 s-\frac{1}{4},2 s+\frac{1}{4}}{\frac{1}{2},s-\frac{1}{8},s+\frac{1}{8},s+\frac{3}{8},s+\frac{5}{8}}{\frac{\lambda }{16}}\\[1.5em]
&\quad-\sum_{s=0}^{\infty} \tfrac{\lambda^{s+1}}{(s+1)!} \;x^{4 (s+1)-2}\;\left(\frac{\Gamma \left(4 (s+1)-\frac{3}{2}\right)}{4 \Gamma \left(4 (s+1)-\frac{1}{2}\right)}\right)\; \left(\tfrac{(4(s+1))!}{(4(s+1) -2)!}\right)\;\cdot\\[1.5em]
&\qquad\pFq{5}{5}{(s+1)+\frac{1}{4},(s+1)+\frac{1}{2},(s+1)+\frac{3}{4},2 (s+1)-\frac{3}{4},2 (s+1)-\frac{1}{4}}{\frac{3}{2},(s+1)-\frac{1}{8},(s+1)+\frac{1}{8},(s+1)+\frac{3}{8},(s+1)+\frac{5}{8}}{\frac{\lambda }{16}}\\[1.5em]
\\
\cG_{5,0}(\lambda;x)&=\sum_{s=0}^{\infty} \tfrac{\lambda^s}{s!} \; x^{5s} \; \pFq{13}{14}{\frac{s}{2}+\frac{1}{10},\frac{s}{2}+\frac{1}{5},\frac{s}{2}+\frac{3}{10},\frac{s}{2}+\frac{2}{5},\frac{s}{2}+\frac{3}{5},\frac{s}{2}+\frac{7}{10},\frac{s}{2}+\frac{4}{5},\frac{s}{2}+\frac{9}{10},s-\frac{1}{10},s+\frac{1}{10},s+\frac{3}{10},s+\frac{1}{2},s+\frac{7}{10}}{\frac{1}{5},\frac{2}{5},\frac{3}{5},\frac{4}{5},\frac{s}{2}-\frac{1}{20},\frac{s}{2}+\frac{1}{20},\frac{s}{2}+\frac{3}{20},\frac{s}{2}+\frac{1}{4},\frac{s}{2}+\frac{7}{20},\frac{s}{2}+\frac{9}{20},\frac{s}{2}+\frac{11}{20},\frac{s}{2}+\frac{13}{20},\frac{s}{2}+\frac{3}{4},\frac{s}{2}+\frac{17}{20}}{-\frac{\lambda ^2}{4096}}\\[1.5em]
&\quad-\sum_{s=0}^{\infty} \tfrac{\lambda^{s+1}}{(s+1)!} \;x^{5 (s+1)-2}\;\left(\frac{\Gamma \left(5 (s+1)-\frac{3}{2}\right)}{4 \Gamma \left(5 (s+1)-\frac{1}{2}\right)}\right)\; \left(\tfrac{(5(s+1))!}{(5(s+1) -2)!}\right)\;\cdot\\[1.5em]
&\qquad\pFq{13}{14}{\frac{s+1}{2}+\frac{1}{10},\frac{s+1}{2}+\frac{1}{5},\frac{s+1}{2}+\frac{3}{10},\frac{s+1}{2}+\frac{2}{5},\frac{s+1}{2}+\frac{3}{5},\frac{s+1}{2}+\frac{7}{10},\frac{s+1}{2}+\frac{4}{5},\frac{s+1}{2}+\frac{9}{10},(s+1)-\frac{3}{10},(s+1)-\frac{1}{10},(s+1)+\frac{1}{10},(s+1)+\frac{3}{10},(s+1)+\frac{1}{2}}{\frac{2}{5},\frac{3}{5},\frac{4}{5},\frac{6}{5},\frac{s+1}{2}-\frac{1}{20},\frac{s+1}{2}+\frac{1}{20},\frac{s+1}{2}+\frac{3}{20},\frac{s+1}{2}+\frac{1}{4},\frac{s+1}{2}+\frac{7}{20},\frac{s+1}{2}+\frac{9}{20},\frac{s+1}{2}+\frac{11}{20},\frac{s+1}{2}+\frac{13}{20},\frac{s+1}{2}+\frac{3}{4},\frac{s+1}{2}+\frac{17}{20}}{-\frac{\lambda ^2}{4096}}\\[1.5em]
&\quad+\sum_{s=0}^{\infty} \tfrac{\lambda^{s+1}}{(s+1)!} \;x^{5 (s+1)-4}\;\left(\frac{\Gamma \left(5 (s+1)-\frac{5}{2}\right)}{16 \Gamma \left(5 (s+1)-\frac{1}{2}\right)}\right)\; \left(\tfrac{(5(s+1))!}{2!(5(s+1) -4)!}\right)\;\cdot\\[1.5em]
&\qquad\pFq{13}{14}{\frac{s+1}{2}+\frac{1}{10},\frac{s+1}{2}+\frac{1}{5},\frac{s+1}{2}+\frac{3}{10},\frac{s+1}{2}+\frac{2}{5},\frac{s+1}{2}+\frac{3}{5},\frac{s+1}{2}+\frac{7}{10},\frac{s+1}{2}+\frac{4}{5},\frac{s+1}{2}+\frac{9}{10},(s+1)-\frac{1}{2},(s+1)-\frac{3}{10},(s+1)-\frac{1}{10},(s+1)+\frac{1}{10},(s+1)+\frac{3}{10}}{\frac{3}{5},\frac{4}{5},\frac{6}{5},\frac{7}{5},\frac{s+1}{2}-\frac{1}{20},\frac{s+1}{2}+\frac{1}{20},\frac{s+1}{2}+\frac{3}{20},\frac{s+1}{2}+\frac{1}{4},\frac{s+1}{2}+\frac{7}{20},\frac{s+1}{2}+\frac{9}{20},\frac{s+1}{2}+\frac{11}{20},\frac{s+1}{2}+\frac{13}{20},\frac{s+1}{2}+\frac{3}{4},\frac{s+1}{2}+\frac{17}{20}}{-\frac{\lambda ^2}{4096}}\\[1.5em]
&\quad-\sum_{s=0}^{\infty} \tfrac{\lambda^{s+2}}{(s+2)!} \;x^{5 (s+2)-6}\;\left(\frac{\Gamma \left(5 (s+2)-\frac{7}{2}\right)}{64 \Gamma \left(5 (s+2)-\frac{1}{2}\right)}\right)\; \left(\tfrac{(5(s+2))!}{3!(5(s+2) -6)!}\right)\;\cdot\\[1.5em]
&\qquad\pFq{13}{14}{\frac{s+2}{2}+\frac{1}{10},\frac{s+2}{2}+\frac{1}{5},\frac{s+2}{2}+\frac{3}{10},\frac{s+2}{2}+\frac{2}{5},\frac{s+2}{2}+\frac{3}{5},\frac{s+2}{2}+\frac{7}{10},\frac{s+2}{2}+\frac{4}{5},\frac{s+2}{2}+\frac{9}{10},(s+2)-\frac{7}{10},(s+2)-\frac{1}{2},(s+2)-\frac{3}{10},(s+2)-\frac{1}{10},(s+2)+\frac{1}{10}}{\frac{4}{5},\frac{6}{5},\frac{7}{5},\frac{8}{5},\frac{s+2}{2}-\frac{1}{20},\frac{s+2}{2}+\frac{1}{20},\frac{s+2}{2}+\frac{3}{20},\frac{s+2}{2}+\frac{1}{4},\frac{s+2}{2}+\frac{7}{20},\frac{s+2}{2}+\frac{9}{20},\frac{s+2}{2}+\frac{11}{20},\frac{s+2}{2}+\frac{13}{20},\frac{s+2}{2}+\frac{3}{4},\frac{s+2}{2}+\frac{17}{20}}{-\frac{\lambda ^2}{4096}}\\[1.5em]
&\quad+\sum_{s=0}^{\infty} \tfrac{\lambda^{s+2}}{(s+2)!} \;x^{5 (s+2)-8}\;\left(\frac{\Gamma \left(5 (s+2)-\frac{9}{2}\right)}{256 \Gamma \left(5 (s+2)-\frac{1}{2}\right)}\right)\; \left(\tfrac{(5(s+2))!}{4!(5(s+2) -8)!}\right)\;\cdot\\[1.5em]
&\qquad\pFq{13}{14}{\frac{s+2}{2}+\frac{1}{10},\frac{s+2}{2}+\frac{1}{5},\frac{s+2}{2}+\frac{3}{10},\frac{s+2}{2}+\frac{2}{5},\frac{s+2}{2}+\frac{3}{5},\frac{s+2}{2}+\frac{7}{10},\frac{s+2}{2}+\frac{4}{5},\frac{s+2}{2}+\frac{9}{10},(s+2)-\frac{9}{10},(s+2)-\frac{7}{10},(s+2)-\frac{1}{2},(s+2)-\frac{3}{10},(s+2)-\frac{1}{10}}{\frac{6}{5},\frac{7}{5},\frac{8}{5},\frac{9}{5},\frac{s+2}{2}-\frac{1}{20},\frac{s+2}{2}+\frac{1}{20},\frac{s+2}{2}+\frac{3}{20},\frac{s+2}{2}+\frac{1}{4},\frac{s+2}{2}+\frac{7}{20},\frac{s+2}{2}+\frac{9}{20},\frac{s+2}{2}+\frac{11}{20},\frac{s+2}{2}+\frac{13}{20},\frac{s+2}{2}+\frac{3}{4},\frac{s+2}{2}+\frac{17}{20}}{-\frac{\lambda ^2}{4096}}\\[1.5em]
\\
\cG_{6,0}(\lambda;x)&=\sum_{s=0}^{\infty} \tfrac{\lambda^s}{s!} \; x^{6s} \; \pFq{8}{8}{s+\frac{1}{6},s+\frac{1}{3},s+\frac{1}{2},s+\frac{2}{3},s+\frac{5}{6},2 s-\frac{1}{6},2 s+\frac{1}{6},2 s+\frac{1}{2}}{\frac{1}{3},\frac{2}{3},s-\frac{1}{12},s+\frac{1}{12},s+\frac{1}{4},s+\frac{5}{12},s+\frac{7}{12},s+\frac{3}{4}}{-\frac{\lambda }{64}}\\[1.5em]
&\quad-\sum_{s=0}^{\infty} \tfrac{\lambda^{s+1}}{(s+1)!} \;x^{6 (s+1)-2}\;\left(\frac{\Gamma \left(6 (s+1)-\frac{3}{2}\right)}{4 \Gamma \left(6 (s+1)-\frac{1}{2}\right)}\right)\; \left(\tfrac{(6(s+1))!}{(6(s+1) -2)!}\right)\;\cdot\\[1.5em]
&\qquad\pFq{8}{8}{(s+1)+\frac{1}{6},(s+1)+\frac{1}{3},(s+1)+\frac{1}{2},(s+1)+\frac{2}{3},(s+1)+\frac{5}{6},2 (s+1)-\frac{1}{2},2 (s+1)-\frac{1}{6},2 (s+1)+\frac{1}{6}}{\frac{2}{3},\frac{4}{3},(s+1)-\frac{1}{12},(s+1)+\frac{1}{12},(s+1)+\frac{1}{4},(s+1)+\frac{5}{12},(s+1)+\frac{7}{12},(s+1)+\frac{3}{4}}{-\frac{\lambda }{64}}\\[1.5em]
&\quad+\sum_{s=0}^{\infty} \tfrac{\lambda^{s+1}}{(s+1)!} \;x^{6 (s+1)-4}\;\left(\frac{\Gamma \left(6 (s+1)-\frac{5}{2}\right)}{16 \Gamma \left(6 (s+1)-\frac{1}{2}\right)}\right)\; \left(\tfrac{(6(s+1))!}{2!(6(s+1) -4)!}\right)\;\cdot\\[1.5em]
&\qquad\pFq{8}{8}{(s+1)+\frac{1}{6},(s+1)+\frac{1}{3},(s+1)+\frac{1}{2},(s+1)+\frac{2}{3},(s+1)+\frac{5}{6},2 (s+1)-\frac{5}{6},2 (s+1)-\frac{1}{2},2 (s+1)-\frac{1}{6}}{\frac{4}{3},\frac{5}{3},(s+1)-\frac{1}{12},(s+1)+\frac{1}{12},(s+1)+\frac{1}{4},(s+1)+\frac{5}{12},(s+1)+\frac{7}{12},(s+1)+\frac{3}{4}}{-\frac{\lambda }{64}}\\[1.5em]
\\
\cG_{8,0}(\lambda;x)&=\sum_{s=0}^{\infty} \tfrac{\lambda^s}{s!} \; x^{8s} \; \pFq{11}{11}{s+\frac{1}{8},s+\frac{1}{4},s+\frac{3}{8},s+\frac{1}{2},s+\frac{5}{8},s+\frac{3}{4},s+\frac{7}{8},2 s-\frac{1}{8},2 s+\frac{1}{8},2 s+\frac{3}{8},2 s+\frac{5}{8}}{\frac{1}{4},\frac{1}{2},\frac{3}{4},s-\frac{1}{16},s+\frac{1}{16},s+\frac{3}{16},s+\frac{5}{16},s+\frac{7}{16},s+\frac{9}{16},s+\frac{11}{16},s+\frac{13}{16}}{\frac{\lambda }{256}}\\[1.5em]
&\quad-\sum_{s=0}^{\infty} \tfrac{\lambda^{s+1}}{(s+1)!} \;x^{8 (s+1)-2}\;\left(\frac{\Gamma \left(8 (s+1)-\frac{3}{2}\right)}{4 \Gamma \left(8 (s+1)-\frac{1}{2}\right)}\right)\; \left(\tfrac{(8(s+1))!}{(8(s+1) -2)!}\right)\;\cdot\\[1.5em]
&\qquad\pFq{11}{11}{(s+1)+\frac{1}{8},(s+1)+\frac{1}{4},(s+1)+\frac{3}{8},(s+1)+\frac{1}{2},(s+1)+\frac{5}{8},(s+1)+\frac{3}{4},(s+1)+\frac{7}{8},2 (s+1)-\frac{3}{8},2 (s+1)-\frac{1}{8},2 (s+1)+\frac{1}{8},2 (s+1)+\frac{3}{8}}{\frac{1}{2},\frac{3}{4},\frac{5}{4},(s+1)-\frac{1}{16},(s+1)+\frac{1}{16},(s+1)+\frac{3}{16},(s+1)+\frac{5}{16},(s+1)+\frac{7}{16},(s+1)+\frac{9}{16},(s+1)+\frac{11}{16},(s+1)+\frac{13}{16}}{\frac{\lambda }{256}}\\[1.5em]
&\quad+\sum_{s=0}^{\infty} \tfrac{\lambda^{s+1}}{(s+1)!} \;x^{8 (s+1)-4}\;\left(\frac{\Gamma \left(8 (s+1)-\frac{5}{2}\right)}{16 \Gamma \left(8 (s+1)-\frac{1}{2}\right)}\right)\; \left(\tfrac{(8(s+1))!}{2!(8(s+1) -4)!}\right)\;\cdot\\[1.5em]
&\qquad\pFq{11}{11}{(s+1)+\frac{1}{8},(s+1)+\frac{1}{4},(s+1)+\frac{3}{8},(s+1)+\frac{1}{2},(s+1)+\frac{5}{8},(s+1)+\frac{3}{4},(s+1)+\frac{7}{8},2 (s+1)-\frac{5}{8},2 (s+1)-\frac{3}{8},2 (s+1)-\frac{1}{8},2 (s+1)+\frac{1}{8}}{\frac{3}{4},\frac{5}{4},\frac{3}{2},(s+1)-\frac{1}{16},(s+1)+\frac{1}{16},(s+1)+\frac{3}{16},(s+1)+\frac{5}{16},(s+1)+\frac{7}{16},(s+1)+\frac{9}{16},(s+1)+\frac{11}{16},(s+1)+\frac{13}{16}}{\frac{\lambda }{256}}\\[1.5em]
&\quad-\sum_{s=0}^{\infty} \tfrac{\lambda^{s+1}}{(s+1)!} \;x^{8 (s+1)-6}\;\left(\frac{\Gamma \left(8 (s+1)-\frac{7}{2}\right)}{64 \Gamma \left(8 (s+1)-\frac{1}{2}\right)}\right)\; \left(\tfrac{(8(s+1))!}{3!(8(s+1) -6)!}\right)\;\cdot\\[1.5em]
&\qquad\pFq{11}{11}{(s+1)+\frac{1}{8},(s+1)+\frac{1}{4},(s+1)+\frac{3}{8},(s+1)+\frac{1}{2},(s+1)+\frac{5}{8},(s+1)+\frac{3}{4},(s+1)+\frac{7}{8},2 (s+1)-\frac{7}{8},2 (s+1)-\frac{5}{8},2 (s+1)-\frac{3}{8},2 (s+1)-\frac{1}{8}}{\frac{5}{4},\frac{3}{2},\frac{7}{4},(s+1)-\frac{1}{16},(s+1)+\frac{1}{16},(s+1)+\frac{3}{16},(s+1)+\frac{5}{16},(s+1)+\frac{7}{16},(s+1)+\frac{9}{16},(s+1)+\frac{11}{16},(s+1)+\frac{13}{16}}{\frac{\lambda }{256}}\\[1.5em]
%\end{tabular}
%\end{table}
\end{longtable}

%%%%%%%%%%%%%%%%%%%%%%%%%%%%%%%%%%%%%%%%%%%%%%%
% change everything back
\newpage
\restoregeometry
\paperwidth=\pdfpageheight
\paperheight=\pdfpagewidth
\pdfpageheight=\paperheight
\pdfpagewidth=\paperwidth
\headwidth=\textwidth
%\end{landscape}
%\normalsize

\section{Proof of the SJ Generating Function Formulae}
\label{app:SJgen}

Recalling the definitions
\begin{equation}
P_n(x,y):=e^{\bB}(xy)^n\,,\quad \bB:=\bfb_0 \partial^2_x\,,\quad \bfb_p:=-\tfrac{1}{2}\tfrac{1}{\hat{D}_x+\hat{D}_y+p-1}\,
\end{equation}
we first prove the following auxiliary set of formulae:
\begin{Lemma}
The differential operators $\bfb_m$ ($m\in \bZ$) fulfill the following relations (for $p,q\in \bZ_{\geq 0}$):
\begin{equation}\label{eq:auxBB}
\begin{aligned}
\bfb_m\hat{y}^p\partial_x^q&=\hat{y}^p\partial_x^q\bfb_{m+p-q}\,,\quad & \hat{y}^p\partial_x^q\bfb_{n}&=\bfb_{m-p+q}\hat{y}^p\partial_x^q\,
\end{aligned}
\end{equation}
and therefore, in particular for all $p\in \bZ_{\geq 0}$,
\begin{equation}\label{eq:BBcomm}
\bfb_m \hat{y}^p\partial_x^p= \hat{y}^p\partial_x^p\bfb_m\,
\end{equation}
\begin{proof}
Via a direct computation, employing the multi-variate analogues of the relations~\eqref{eq:ENO1} presented in Lemma~\ref{lem:ENO}.
\end{proof}
\end{Lemma}
Next, we prove by induction the following formula:
\begin{equation}
\bB^m e^{\lambda xy}=(\lambda \hat{y})^{2m}\left(\prod_{j=0}^{m-1}\bfb_{2(m+j)}\right)e^{\lambda xy}\,
\end{equation}

Starting the proof with verifying the case $m=1$  (making  use of the auxiliary relations~\eqref{eq:auxBB}),
\begin{align*}
\bB e^{\lambda xy}&=\bfb_0 \partial_x^2 e^{\lambda xy}
=\lambda^2 \bfb_0 \hat{y}^2 e^{\lambda xy}
\overset{\eqref{eq:auxBB}}{=}
(\lambda \hat{y})^2 \bfb_2 e^{\lambda xy}\,
\end{align*}%
the induction step $m\to m+1$ may be verified as follows:
{\allowdisplaybreaks
\begin{align*}
\bB\left(\bB^m e^{\lambda xy}\right)
&=\bfb_0\partial_x^2 (\lambda \hat{y})^{2m}\left(\prod_{j=0}^{m-1}\bfb_{2(m+j)}\right)e^{\lambda xy}\\
&\overset{\eqref{eq:BBcomm}}{=}\lambda^{2m}\bfb_{0}\hat{y}^{2(m-1)}
\left(\prod_{j=0}^{m-1}\bfb_{2(m+j)}\right)\hat{y}^2\partial_x^2e^{\lambda xy}\\
&=\lambda^{2(m+1)}\hat{y}^{2(m-1)}\bfb_{2(m-1)}
\left(\prod_{j=0}^{m-1}\bfb_{2(m+j)}\right)\hat{y}^4e^{\lambda xy}\\
&=(\lambda \hat{y})^{2(m+1)}\bfb_{2(m+1)}
\left(\prod_{j=0}^{m-1}\bfb_{2((m+1)+j+1)}\right)e^{\lambda xy}\\
&=(\lambda \hat{y})^{2(m+1)}
\left(\prod_{j=0}^{(m+1)-1}\bfb_{2((m+1)+j)}\right)e^{\lambda xy}\,
\end{align*}}
which proves the induction hypothesis.

Assembling all the auxiliary formulae, it is then straightforward to prove the explicit formula~\eqref{eq:SJegf1} for the EGF $\cG(\lambda;x,y)$ of the polynomials $P_n(x,y)$:
{\allowdisplaybreaks
\begin{align*}
\cG(\lambda;x,y)&=e^{\bB}e^{\lambda xy}=e^{\lambda xy}
+\sum_{m\geq 1}\frac{(\lambda^2 \hat{y}^2)^m}{m!}\left(\prod_{j=0}^{m-1}\bfb_{2(m+j)}\right)e^{\lambda xy}\\
&=\sum_{n\geq 0}\frac{1}{n!}\left[1+\sum_{m\geq 1}\frac{(\lambda^2 \hat{y}^2)^m}{m!}\left(\prod_{j=0}^{m-1}\bfb_{2(m+j)}\right)\right](xy)^n\\
&=\sum_{n\geq 0}\frac{(\lambda x y)^n}{n!}\left[
1
+\sum_{m\geq 1}\tfrac{1}{m!}\left(-\tfrac{\lambda^2 y^2}{2}\right)^m \left(\prod_{j=0}^{m-1}\tfrac{1}{2(n+m+j)-1}\right)
\right]\\
&=\sum_{n\geq 0}\frac{(\lambda x y)^n}{n!}\left[
1
+\sum_{m\geq 1}\tfrac{1}{m!}\left(-\tfrac{\lambda^2 y^2}{4}\right)^m \tfrac{\Gamma(m+n-\tfrac{1}{2})}{\Gamma(2m+n-\tfrac{1}{2})}
\right]\\
&=\sum_{n\geq 0}\frac{(\lambda x y)^n}{n!}\sum_{m\geq 0}\tfrac{1}{m!}\left(-\tfrac{\lambda^2 y^2}{4}\right)^m \tfrac{\Gamma(m+n-\tfrac{1}{2})}{\Gamma(2m+n-\tfrac{1}{2})}\,
\end{align*}}

Finally, invoking the auxiliary formula~\eqref{eq:GMFC} to compute
\begin{equation}
\Gamma(2m+n-\tfrac{1}{2})=4^m \Gamma(n-\tfrac{1}{2})\left(\tfrac{n}{2}-\tfrac{1}{4}\right)_m\left(\tfrac{n}{2}+\tfrac{1}{4}\right)_m
\end{equation}
and since
\begin{equation}
\Gamma(m+n-\tfrac{1}{2})=\left(n-\tfrac{1}{2}\right)_m\;\Gamma(n-\tfrac{1}{2})\,
\end{equation}
we arrive at the final form for $\cG(\lambda;x,y)$:
\begin{equation}
\begin{aligned}
\cG(\lambda;x,y)&=
\sum_{n\geq 0}\frac{(\lambda x y)^n}{n!}
\sum_{m\geq 0}\tfrac{1}{m!}\left(-\tfrac{\lambda^2 y^2}{4}\right)^m \frac{\left(n-\tfrac{1}{2}\right)_m\;\Gamma(n-\tfrac{1}{2})}{%
4^m \Gamma(\tfrac{n}{2}-\tfrac{1}{2})\left(\tfrac{n}{2}-\tfrac{1}{4}\right)_m\left(n+\tfrac{1}{4}\right)_m}\\
&=\sum_{n\geq 0}\frac{(\lambda x y)^n}{n!}\; \pFq{1}{2}{n-\tfrac{1}{2}}{\tfrac{n}{2}-\tfrac{1}{4},\tfrac{n}{2}+\tfrac{1}{4}}{-\tfrac{\lambda^2y^2}{16}}\,
\end{aligned}
\end{equation}

This concludes the proof of the explicit formula for $\cG(\lambda;x,y)$ as given in~\eqref{eq:SJegf1}.

\section{Lacunary Generating Functions of the Two-Variable Hermite Polynomials}
\label{app:HPlgf}

\begin{Theorem}[Prop.~1 and Thm.~1 of~\cite{bdp2018hp}]\label{thm:lacHP}
For $K\in\bZ_{\geq 1}$ and $L\in \bZ_{\geq 0}$, let the EGF of the lacunary shifts $\cH_{K,L}(\lambda;x,z)$ of the $K$-tuple lacunary generating function $\cH_{K,0}(\lambda;x,y)$ of the polynomials $H_n(x,z)$, be defined~as
\begin{equation}
\cH_K(\mu;\lambda;x,z):= \sum_{L\geq 0}\frac{\mu^L}{L!} \cH_{K,L}(\lambda;x,z)
=\bL_K\left(e^{\mu \frac{\partial}{\partial \lambda}}\cH_{1,0}(\lambda;x,z)\right)\,
\end{equation}

Then the explicit formula for $\cH_K(\mu;\lambda;x,z)$ is given by
\begin{equation}\label{eq:propHLGF}
\cH_K(\mu;\lambda;x,z)=e^{\mu x+\mu^2 z}\cH_{K,0}(\lambda;x+2\mu z,z)\,
\end{equation}
where $\cH_{K,0}(\lambda;x,z)$ for $K=2T$ ($T\in \bZ_{\geq 1}$) reads
\begin{equation}\label{eq:lacHPeven}
\begin{aligned}
\cH_{K=2T,0}(\lambda;x,y)
&=\sum_{\beta=0}^{T-1}\sum_{s\geq 0}\frac{\lambda^s}{s!}\; x^{K\cdot s -2\beta}y^{\beta} \; \tilde{h}_{K\cdot s,\beta}\;\cdot\\
&\qquad\pFq{(K-1)}{(T-1)}{\left(s+\tfrac{j+1}{K}\right)_{0\leq j\leq K-2}}{\left(\tfrac{\beta+\ell+1}{T}\right)_{\stackrel{0\leq \ell\leq T-1}{\ell\neq T-1-\beta}}}{\lambda(2y K)^T}\,
\end{aligned}
\end{equation}
while for $K=2T+1$ (with $T\in \bZ_{\geq1}$)
\begin{equation}\label{eq:lacHPodd}
\begin{aligned}
\cH_{K=2T+1,0}(\lambda;x,y)&=\sum_{\beta=0}^{K-1}\sum_{s\geq 0}\frac{\lambda^s}{s!}\; x^{K\cdot s-2\beta}y^{\beta}\; \tilde{h}_{K\cdot s,\beta}\;\cdot\\
&\qquad
\pFq{(2K-2)}{(K-1)}{\left(\tfrac{s}{2}+\tfrac{j+1}{2K}\right)_{\stackrel{0\leq j\leq 2K-2}{j\neq K-1}}}{\left(\tfrac{\beta+\ell+1}{K}\right)_{\stackrel{0\leq \ell\leq K-1}{\ell\neq K-1-\beta}}}{\tfrac{\lambda^2(4yK)^K}{4}}\,
\end{aligned}
\end{equation}
Here, $\tilde{h}_{n,k}$ denote the so-called \emph{matching coefficients} (of directed Hermite-configurations),
\begin{equation}
\tilde{h}_{n,k}:=\begin{cases}
\frac{n!}{(n-2k)!k!} &\qquad \text{if $0\leq 2k\leq n$}\\
0 &\qquad \text{otherwise}
\end{cases}
\end{equation}
\end{Theorem}

\section{Proof of the Connection Coefficient Formulae}
\label{app:AB}

The generating functions $A(\lambda,\mu)$ and $B(\lambda,\mu)$ verify
\begin{equation*}
\frac{1}{\pi}\int d^2w\; e^{-|w|^2} A(\alpha,w)B(\bar{w},\beta)=e^{\alpha\beta}\,
\end{equation*}

\begin{proof}
Writing out the definitions of $A(\alpha,w)$ and $B(\bar{w},\beta)$ explicitly as
\begin{align*}
A(\alpha,w)
&=\tio\left(
\sqrt{u_1v_1}\; e^{\alpha w u_1v_1+\alpha^2 \tfrac{v_1}{4}}
\right)\\
B(\bar{w},\beta)&=\tio\left(
\frac{1}{\sqrt{u_2v_2}}e^{\beta\bar{w}  u_2v_2+\bar{w}^2\left(-\tfrac{u_2v_2^2}{4}\right)}\right)\,
\end{align*}%
taking care in particular that the respective integration variables $u_1,u_2\in\cU$ and $v_1,v_2\in \cV$ are distinct. Inserting these formulae into the integral in question, we obtain:
\begin{myequation}
\frac{1}{\pi}\int d^2w\; e^{-|w|^2} A(\alpha,w)B(\bar{w},\beta)
=\tio\left(e^{\alpha^2 \tfrac{v_1}{4}}\;\sqrt{\frac{u_1v_1}{u_2 v_2}}\;
\frac{1}{\pi}\int d^2w\; e^{-|w|^2}
\; e^{ w \alpha u_1v_1+\bar{w}  \beta u_2v_2+\bar{w}^2\left(-\tfrac{u_2v_2^2}{4}\right)}
\right)
\,
\end{myequation}%

Making use of the auxiliary integral formula (which itself is a straightforward consequence of the formula for the exponential generating function~\eqref{eq:HGF} of the Hermite polynomials $H_n(x,z)$),
\begin{equation}
\tfrac{1}{\pi}\int d^2\;e^{-|w|^2} e^{ w \mu+ \bar{w} \nu+\bar{w}^2\rho }=e^{\mu \nu+\mu^2\rho}\,
\end{equation}
the proof of the claim thus reduces to verifying the following equation:
\begin{equation}
\begin{aligned}
\frac{1}{\pi}\int d^2w\; e^{-|w|^2} A(\alpha,w)B(\bar{w},\beta)
&=\tio\left(e^{\alpha^2 \tfrac{v_1}{4}}\;\sqrt{\frac{u_1v_1}{u_2 v_2}}
\; e^{(\alpha u_1v_1u_2v_2)\beta+(\alpha u_1v_1u_2v_2)^2\left(-\tfrac{1}{4u_2}\right)}
\right)\\
&=\tio\left(\sqrt{\frac{u_1v_1}{u_2 v_2}}
\; e^{(\alpha u_1v_1u_2v_2)\beta+ \tfrac{\alpha^2v_1}{4}\left(1- u_1^2v_1u_2v_2^2\right)}
\right)
\,
\end{aligned}
\end{equation}%

We complete the proof by providing the following auxiliary formula (for all $N\in \bR\setminus \{0\}$ and for all $p\in\bZ_{\geq 1}$):
\begin{align*}
&
\tio\left(
(u_1v_1)^{N+1}
(u_2v_2)^{N}v_1^p
(1-u_1^2v_1u_2v_2^2)^p\right)=
\tio\left(
(u_1v_1)^{N+1}
(u_2v_2)^{N}v_1^p
\sum_{k=0}^p\binom{p}{k}(-1)^k (u_1^2v_1u_2v_2^2)^k
\right)\\
&\quad=
\tio\left(
\sum_{k=0}^p\binom{p}{k}(-1)^k
{u_1^{N+2k+1}}v_1^{N+k+p+1}u_2^{N+k}{v_2^{N+2k}}
\right)\\
&\quad=
\tio\left(
\sum_{k=0}^p\binom{p}{k}(-1)^k {\frac{\Gamma(N+2k+1)}{\Gamma(N+2k)}}
v_1^{N+k+p+1}u_2^{N+k}
\right)\\
&\quad=
{N}\,
\tio\left(
\sum_{k=0}^p\binom{p}{k}(-1)^k
v_1^{N+k+p+1}u_2^{N+k}
\right)
+{2}\,
\tio\left(
\sum_{k=0}^p\binom{p}{k}(-1)^k {k}
v_1^{N+k+p+1}u_2^{N+k}
\right)\\
&=N\tio\left(
\left(\sum_{k=0}^p\binom{p}{k}(-1)^k e^{k\partial_N}\right)u_2^{N}v_1^{N+p+1}
\right)
-2p\tio\left(
\left(\sum_{k=0}^{p-1}\binom{p-1}{k}(-1)^k e^{k\partial_N}\right)u_2^{N+1}v_1^{N+p+2}
\right)\\
&=N\tio\left(
u_2^{N}v_1^{N+p+1}\left(1-u_2v_1\right)^p
\right)
-2p\tio\left(
\left(1-u_2v_1\right)^{p-1}u_2^{N+1}v_1^{N+p+2}
\right)\,
\end{align*}%

To complete the calculation, we insert $1$ into both terms in the form of monomials $u_3^qv_3^q$ suitable in order to ``complete'' the terms in $(\dotsc)$ into beta functions (compare~\eqref{eq:betaDef}, \eqref{eq:betaprop} and~\eqref{eq:betaProp2}):
\begin{align*}
&N\tio\left(
u_2^{N}v_1^{N+p+1}\left(1-u_2v_1\right)^p
\right)
-2p\tio\left(
\left(1-u_2v_1\right)^{p-1}u_2^{N+1}v_1^{N+p+2}
\right)\\
&\quad=
N\tio\left(
v_3^{p+1}u_3^{p+1}u_2^{N}v_1^{N+p+1}\left(1-u_2v_1\right)^p
\right)
-2p\tio\left(v_3^{p+1}u_3^{p+1}
\left(1-u_2v_1\right)^{p-1}u_2^{N+1}v_1^{N+p+2}
\right)\\
&\quad
=\frac{N}{p!}\left(1-e^{\partial_N}\right)^p
B(N,p+1)
-\frac{2p}{p!}
\left(1-e^{\partial_N}\right)^{p-1}B(N+1,p+1)\\
&\hspace{10pt}=\frac{N}{p!}B(N,2p+1)
-\frac{2p}{p!}B(N+1,2p)\\
&\quad=\frac{1}{p! \Gamma(N+2p+1)}\left(
N\Gamma(N)\Gamma(2p+1)-(2p)\Gamma(N+1)\Gamma(2p)
\right)=0\,
\end{align*}%
Together with the evident identity
\begin{equation}
\tio\left(
(u_1v_1)^{N+1}
(u_2v_2)^{N}\right)=1\,
\end{equation}
this concludes the proof.
\end{proof}

\reftitle{References}


\begin{thebibliography}{999}

\bibitem[Roman and Rota(1978)]{roman1978umbral}
Roman, S.M.; Rota, G.C.
\newblock The umbral calculus.
\newblock {\em Adv. Math.} {\bf 1978}, {\em 27},~95--188. [\href{http://dx.doi.org/10.1016/0001-8708(78)90087-7}{CrossRef}]

\bibitem[Dattoli(2000)]{dattoli2000generalized}
Dattoli, G.
\newblock Generalized polynomials, operational identities and their
applications.
\newblock {\em J. Comput. Appl. Math.} {\bf 2000},
{\em 118},~111--123. [\href{http://dx.doi.org/10.1016/S0377-0427(00)00283-1}{CrossRef}]

\bibitem[Roman(1984)]{roman1984umbral}
Roman, S.~\newblock {\em The Umbral Calculus};  Elsevier: Amsterdam, The Netherlands, 1984; Volume 111. [\href{http://dx.doi.org/10.1016/S0079-8169(08)61589-5}{CrossRef}]

\bibitem[Dattoli \em{et~al.}(1997)Dattoli, Ottaviani, Torre, and
V{\'{a}}zquez]{dattoli1997evolution}
Dattoli, G.; Ottaviani, P.L.; Torre, A.; V{\'{a}}zquez, L.
\newblock {Evolution operator equations: Integration with algebraic and
finite-difference methods. Applications to physical problems in classical and
quantum mechanics and quantum field theory}.
\newblock {\em Rivista Nuovo Cimento} {\bf 1997}, {\em 20},~3--133. [\href{http://dx.doi.org/10.1007/bf02907529}{CrossRef}]

\bibitem[Dattoli \em{et~al.}(1999)Dattoli, Torre, and
Lorenzutta]{dattoli1999operational}
Dattoli, G.; Torre, A.; Lorenzutta, S.
\newblock Operational Identities and Properties of Ordinary and Generalized
Special Functions.
\newblock {\em J. Math. Anal. Appl.} {\bf 1999},
{\em 236},~399--414. [\href{http://dx.doi.org/10.1006/jmaa.1999.6447}{CrossRef}]

\bibitem[Dattoli and Zhukovsky(2007)]{rA-Dattoli:2007}
Dattoli, G.; Zhukovsky, K.V.
\newblock {Quark flavour mixing and the exponential form of the
Kobayashi--Maskawa matrix}.
\newblock {\em Eur. Phys. J. C} {\bf 2007}, {\em 50},~817--821. [\href{http://dx.doi.org/10.1140/epjc/s10052-007-0263-1}{CrossRef}]

\bibitem[Zhukovsky and Dattoli(2011)]{rB-Zhukovsky:2011}
Zhukovsky, K.V.; Dattoli, G.
\newblock {Evolution of non-spreading Airy wavepackets in time dependent linear
potentials}.
\newblock {\em Appl. Math. Comput.} {\bf 2011}, {\em
217},~7966--7974. [\href{http://dx.doi.org/10.1016/j.amc.2011.02.088}{CrossRef}]

\bibitem[Artioli \em{et~al.}(2017)Artioli, Dattoli, Licciardi, and
Pagnutti]{rC-Artioli:2017}
Artioli, M.; Dattoli, G.; Licciardi, S.; Pagnutti, S.
\newblock {Fractional Derivatives, Memory Kernels and Solution of a Free
Electron Laser Volterra Type Equation}.
\newblock {\em Mathematics} {\bf 2017}, {\em 5},~73. [\href{http://dx.doi.org/10.3390/math5040073}{CrossRef}]

\bibitem[Dattoli \em{et~al.}(2017)Dattoli, G{\'o}rska, Horzela, Penson, and
Sabia]{rD-Dattoli:2017}
Dattoli, G.; G{\'o}rska, K.; Horzela, A.; Penson, K.A.; Sabia, E.
\newblock {Theory of relativistic heat polynomials and one-sided L{\'e}vy
distributions}.
\newblock {\em J. Math. Phys.} {\bf 2017}, {\em 58},~063510. [\href{http://dx.doi.org/10.1063/1.4985072}{CrossRef}]


\bibitem[Zhukovsky(2017)]{rE-Zhukovsky:2017}
Zhukovsky, K.V.
\newblock {Operational solution for some types of second order differential
equations and for relevant physical problems}.
\newblock {\em J. Math. Anal. Appl.} {\bf 2017},
{\em 446},~628--647. [\href{http://dx.doi.org/10.1016/j.jmaa.2016.08.054}{CrossRef}]

\bibitem[Zhukovsky(2016{\natexlab{a}})]{rF-Zhukovsky:2016}
Zhukovsky, K.V.
\newblock {The operational solution of fractional-order differential equations,
as well as Black---Scholes and heat-conduction equations}.
\newblock {\em Moscow Univ. Phys. Bull.} {\bf 2016}, {\em
71},~237--244. [\href{http://dx.doi.org/10.3103/S0027134916030164}{CrossRef}]


\bibitem[Zhukovsky(2016{\natexlab{b}})]{rG-Zhukovsky:2016}
Zhukovsky, K.V.
\newblock {Operational method of solution of linear non-integer ordinary and
partial differential equations}.
\newblock {\em SpringerPlus} {\bf 2016}, {\em 5}, 119. [\href{http://dx.doi.org/10.1186/s40064-016-1734-3}{CrossRef}] [\href{http://www.ncbi.nlm.nih.gov/pubmed/26900541}{PubMed}]

\bibitem[Zhukovsky(2017)]{rH-Zhukovsky:2017}
Zhukovsky, K.V.
\newblock {Solving evolutionary-type differential equations and physical
problems using the operator method}.
\newblock {\em Theor. Math. Phys.} {\bf 2017}, {\em
190},~52--68. [\href{http://dx.doi.org/10.1134/s0040577917010044}{CrossRef}]

\bibitem[Zhukovsky(2016)]{rI-Zhukovsky:2016}
Zhukovsky, K.V.
\newblock {Operational Approach and Solutions of Hyperbolic Heat Conduction
Equations}.
\newblock {\em Axioms} {\bf 2016}, {\em 5},~28. [\href{http://dx.doi.org/10.3390/axioms5040028}{CrossRef}]

\bibitem[Dattoli \em{et~al.}(2007)Dattoli, Srivastava, and
Zhukovsky]{rJ-Dattoli:2007}
Dattoli, G.; Srivastava, H.M.; Zhukovsky, K.V.
\newblock {Operational methods and differential equations with applications to
initial-value problems}.
\newblock {\em Appl. Math. Comput.} {\bf 2007}, {\em
184},~979--1001. [\href{http://dx.doi.org/10.1016/j.amc.2006.07.001}{CrossRef}]

\bibitem[Gurappa \em{et~al.}(1996)Gurappa, Kumar, and
Panigrahi]{gurappa1996new}
Gurappa, N.; Kumar, C.N.; Panigrahi, P.K.
\newblock {New exactly and conditionally exactly solvable N-body problems in
one dimension}.
\newblock {\em Modern Phys. Lett. A} {\bf 1996}, {\em 11},~1737--1744. [\href{http://dx.doi.org/10.1142/S0217732396001727}{CrossRef}]


\bibitem[Gurappa \em{et~al.}(2001)Gurappa, Panigrahi, Shreecharan, and
Ranjani]{gurappa2001novel}
Gurappa, N.; Panigrahi, P.K.; Shreecharan, T.; Ranjani, S.S.
\newblock A Novel Method to Solve Familiar Differential Equations and its
Applications. In {\em Frontiers of Fundamental Physics 4}; Springer: New York, NY, USA,
2001; pp.~269--277. [\href{http://dx.doi.org/10.1007/978-1-4615-1339-1\_26}{CrossRef}]

\bibitem[Gurappa \em{et~al.}(2002)Gurappa, Panigrahi, and
Shreecharan]{gurappa2002linear}
Gurappa, N.; Panigrahi, P.K.; Shreecharan, T.
\newblock Linear differential equations and orthogonal polynomials: A~novel
approach.
\newblock \emph{arXiv} {\bf 2002},  arXiv:math-ph/0203015.


\bibitem[Gurappa \em{et~al.}(2003)Gurappa, Panigrahi, and
Shreecharan]{Gurappa_2003}
Gurappa, N.; Panigrahi, P.K.; Shreecharan, T.
\newblock A new perspective on single and multi-variate differential equations.
\newblock {\em J. Comput. Appl. Math.} {\bf 2003},
{\em 160},~103--112. [\href{http://dx.doi.org/10.1016/S0377-0427(03)00616-2}{CrossRef}]

\bibitem[Babusci \em{et~al.}(2013)Babusci, Dattoli, G{\'{o}}rska, and
Penson]{babusci2013symbolic}
Babusci, D.; Dattoli, G.; G{\'{o}}rska, K.; Penson, K.A.
\newblock {Symbolic methods for the evaluation of sum rules of Bessel
functions}.
\newblock {\em J. Math. Phys.} {\bf 2013}, {\em 54},~073501. [\href{http://dx.doi.org/10.1063/1.4812325}{CrossRef}]

\bibitem[Babusci \em{et~al.}(2014)Babusci, Dattoli, G{\'{o}}rska, and
Penson]{babusci2014spherical}
Babusci, D.; Dattoli, G.; G{\'{o}}rska, K.; Penson, K.
\newblock {The spherical Bessel and Struve functions and operational methods}.
\newblock {\em Appl. Math. Comput.} {\bf 2014}, {\em
238},~1--6. [\href{http://dx.doi.org/10.1016/j.amc.2014.03.137}{CrossRef}]

\bibitem[Dattoli \em{et~al.}(2017)Dattoli, di~Palma, Sabia, G{\'{o}}rska,
Horzela, and Penson]{dattoli2017operational}
Dattoli, G.; di~Palma, E.; Sabia, E.; G{\'{o}}rska, K.; Horzela, A.; Penson,
K.A.
\newblock {Operational Versus Umbral Methods and the Borel Transform}.
\newblock {\em Int. J. Appl. Comput. Math.}
{\bf 2017}, {\em 3},~3489--3510. [\href{http://dx.doi.org/10.1007/s40819-017-0315-7}{CrossRef}]


\bibitem[Licciardi(2018)]{licciardi2018umbral}
Licciardi, S.
\newblock {Umbral Calculus, a Different Mathematical Language}. Ph.D. Thesis,  University of Catania,  Catania, Italy,
\newblock 2018.

\bibitem[Behr \em{et~al.}(2017)Behr, Duchamp, and Penson]{bdp2017}
Behr, N.; Duchamp, G.H.E.; Penson, K.A.
\newblock {Combinatorics of chemical reaction systems}.
\newblock \emph{arXiv} {\bf 2017}, arXiv:1712.06575.

\bibitem[G{\'{o}}rska \em{et~al.}(2012)G{\'{o}}rska, Babusci, Dattoli, Duchamp,
and Penson]{babusci2011ramanujan}
G{\'{o}}rska, K.; Babusci, D.; Dattoli, G.; Duchamp, G.; Penson, K.
\newblock {The Ramanujan master theorem and its implications for special
functions}.
\newblock {\em Appl. Math. Comput.} {\bf 2012}, {\em
218},~11466--11471. [\href{http://dx.doi.org/10.1016/j.amc.2012.05.036}{CrossRef}]

\bibitem[{\relax DLMF}()]{DLMF}
Olver, F.W.J.; {Olde Daalhuis}, A.B.; Lozier, D.W.;  Schneider, B.I.;
Boisvert, R.F.; Clark, C.W.; Miller,~B.R.; Saunders, B.V. (Eds.)
{NIST Digital Library of Mathematical Functions}.
Available online: \newblock \href{http://dlmf.nist.gov/}{http://dlmf.nist.gov/}(accessed on 27 March 2018).


\bibitem[Brychkov(2008)]{brychkov2008handbook}
Brychkov, Y.A.
\newblock {\em {Handbook of Special Functions: Derivatives, Integrals, Series
and Other Formulas}}; Chapman and Hall/{CRC}: London, UK,  2008. [\href{http://dx.doi.org/10.1201/9781584889571}{CrossRef}]

\bibitem[Kwon \em{et~al.}(1994)Kwon, Littlejohn, and
Yoo]{kwon1994characterizations}
Kwon, K.H.; Littlejohn, L.L.; Yoo, B.H.
\newblock {Characterizations of Orthogonal Polynomials Satisfying Differential
Equations}.
\newblock {\em {SIAM} J. Math. Anal.} {\bf 1994}, {\em
25},~976--990. [\href{http://dx.doi.org/10.1137/S0036141092236437}{CrossRef}]

\bibitem[Kwon \em{et~al.}(1996)Kwon, Littlejohn, and Yoo]{kwon1996new}
Kwon, K.; Littlejohn, L.; Yoo, B.
\newblock New characterizations of classical orthogonal polynomials.
\newblock {\em Indag. Math.} {\bf 1996}, {\em 7},~199--213. [\href{http://dx.doi.org/10.1016/0019-3577(96)85090-7}{CrossRef}]

\bibitem[Kwon and Littlejohn(1998)]{kwon1998sobolev}
Kwon, K.; Littlejohn, L.
\newblock {Sobolev Orthogonal Polynomials and Second-Order Differential
Equations}.
\newblock {\em Rocky~Mt. J.~Math.} {\bf 1998}, {\em
28},~547--594. [\href{http://dx.doi.org/10.1216/rmjm/1181071786}{CrossRef}]

\bibitem[Olver \em{et~al.}(2010)Olver, Lozier, Boisvert, and
Clark]{olver2010nist}
Olver, F.W.; Lozier, D.W.; Boisvert, R.F.; Clark, C.W.
\newblock {\em {NIST Handbook of Mathematical Functions}}, 1st ed.; Cambridge
University Press: New York, NY, USA,  2010.



\bibitem[Bruder and Littlejohn(2012)]{Bruder_2012}
Bruder, A.; Littlejohn, L.L.
\newblock {Classical and Sobolev orthogonality of the nonclassical Jacobi
polynomials with parameters {$\alpha=\beta=-1$}}.
\newblock {\em Annali di Matematica Pura ed Applicata} {\bf 2012}, {\em
193},~431--455. [\href{http://dx.doi.org/10.1007/s10231-012-0284-8}{CrossRef}]


\bibitem[Gurappa and Panigrahi(1999)]{gurappa1999equivalence}
Gurappa, N.; Panigrahi, P.K.
\newblock Equivalence of the Sutherland model to free particles on a circle.
\newblock \emph{arXiv} {\bf 1999}, arXiv:9908127.

\bibitem[Shreecharan \em{et~al.}(2004)Shreecharan, Panigrahi, and
Banerji]{Shreecharan_2004}
Shreecharan, T.; Panigrahi, P.K.; Banerji, J.
\newblock Coherent states for exactly solvable potentials.
\newblock {\em Phys. Rev. A} {\bf 2004}, {\em 69}, 012102. [\href{http://dx.doi.org/10.1103/PhysRevA.69.012102}{CrossRef}]

\bibitem[Gurappa and Panigrahi(2004)]{gurappa2004polynomial}
Gurappa, N.; Panigrahi, P.K.
\newblock On polynomial solutions of the Heun equation.
\newblock {\em J. Phys. A} {\bf 2004}, {\em
37},~L605--L608. [\href{http://dx.doi.org/10.1088/0305-4470/37/46/L01}{CrossRef}]

\bibitem[Gurappa(2007)]{gurappa2007applications}
Gurappa, N.
\newblock {On the Applications of a New Technique to Solve Linear Differential
Equations, with and without Source}.
\newblock {\em Symmetry, Integr. Geom. Methods Appl.}
{\bf 2007}. [\href{http://dx.doi.org/10.3842/SIGMA.2007.057}{CrossRef}]

\bibitem[Dattoli and Sabia(2010)]{dattoli2010generalized}
Dattoli, G.; Sabia, E.
\newblock {Generalized Transforms and Special Functions, RT/2009/ENEA}.
\newblock \emph{arXiv} {\bf 2010}, arXiv:1010.1679.


\bibitem[Srivastava and Manocha()]{srivastavatreatise}
Srivastava, H.; Manocha, H.
\newblock \emph{A Treatise on Generating Functions};
\newblock Ellis Harwood Limited: Chichester, West Sussex, UK, 1984.

\bibitem[Dattoli \em{et~al.}(2000)Dattoli, Lorenzutta, and
Sacchetti]{dattoli2000note}
Dattoli, G.; Lorenzutta, S.; Sacchetti, D.
\newblock \emph{A Note on Operational Rules for Hermite and Laguerre Polynomials};
\newblock Technical Report; ENEA Frascati: Rome, Italy,  2000.

\bibitem[Wilf(2005)]{wilf2005generatingfunctionology}
Wilf, H.
\newblock {\em Generatingfunctionology}; A K Peters/{CRC} Press: New York, NY, USA, 2005. [\href{http://dx.doi.org/10.1201/b10576}{CrossRef}]

\bibitem[Behr \em{et~al.}(2018)Behr, Duchamp, and Penson]{bdp2018hp}
Behr, N.; Duchamp, G.H.E.; Penson, K.A.
\newblock {Explicit formulae for all higher order exponential lacunary
generating functions of Hermite polynomials}. \emph{arXiv} {\bf 2018}, arXiv:1806.08417.

\end{thebibliography}
\end{document}